\documentclass[pra,twocolumn,showpacs,groupedaddress,superscriptaddress,nofootinbib,floatfix,preprintnumbers,longbibliography]{revtex4-1}

\usepackage{amssymb,amsmath,graphicx,color}
\usepackage{bm}
\usepackage[tight]{subfigure}
\usepackage[export]{adjustbox}
\usepackage{braket}
\usepackage[colorlinks=true,citecolor=blue,linkcolor=red]{hyperref}

\begin{document}

\title{
Towards analog quantum simulations of lattice gauge theories with trapped ions
}

\author{Zohreh Davoudi}
\thanks{The authors' list is alphabetically ordered.}
\affiliation{
Maryland Center for Fundamental Physics and Department of Physics, 
University of Maryland, College Park, MD 20742, USA}
\affiliation{
RIKEN Center for Accelerator-based Sciences,
Wako 351-0198, Japan}

\author{Mohammad Hafezi}
\affiliation{
Joint Quantum Institute and Department of Physics,
University of Maryland, College Park, MD 20742}
\affiliation{
Department of Electrical and Computer Engineering and \\
Institute for Research in Electronics and Applied Physics, University of Maryland, College Park, MD 20742}

\author{Christopher Monroe}
\affiliation{
Joint Quantum Institute and Department of Physics,
University of Maryland, College Park, MD 20742}
\affiliation{Joint Center for Quantum Information and Computer Science, University of Maryland, College Park, MD 20742, USA}

\author{Guido Pagano}
\affiliation{
Joint Quantum Institute and Department of Physics,
University of Maryland, College Park, MD 20742}
\affiliation{Joint Center for Quantum Information and Computer Science, University of Maryland, College Park, MD 20742, USA}

\author{Alireza Seif}
\affiliation{
Joint Quantum Institute and Department of Physics,
University of Maryland, College Park, MD 20742}

\author{Andrew Shaw}
\affiliation{
Maryland Center for Fundamental Physics and Department of Physics, 
University of Maryland, College Park, MD 20742, USA}

\date{\today}

\preprint{UMD-PP-019-03}


\begin{abstract}
Gauge field theories play a central role in modern physics and are at the heart of the Standard Model of elementary particles and interactions. Despite significant progress in applying classical computational techniques to simulate gauge theories, it has remained a challenging task to compute the real-time dynamics of systems described by gauge theories. An exciting possibility that has been explored in recent years is the use of highly-controlled quantum systems to simulate, in an analog fashion, properties of a target system whose dynamics are difficult to compute. Engineered atom-laser interactions in a linear crystal of trapped ions offer a wide range of possibilities for quantum simulations of complex physical systems. Here, we devise practical proposals for analog simulation of simple lattice gauge theories whose dynamics can be mapped onto spin-spin interactions in any dimension. These include 1+1D quantum electrodynamics, 2+1D Abelian Chern-Simons theory coupled to fermions, and 2+1D pure $Z_2$ gauge theory. The scheme proposed, along with the optimization protocol applied, will have applications beyond the examples presented in this work, and will enable scalable analog quantum simulation of Heisenberg spin models in any number of dimensions and with arbitrary interaction strengths.
\end{abstract}

\maketitle
\section{Introduction
\label{sec:intro}}
\noindent
The invariance of physical systems under local transformations of fields leads to fundamental constraints on how matter fields interact, and introduces new bosonic degrees of freedom, the gauge fields. Gauge field theories coupled to matter are responsible for a wide range of phenomena in nature, and permeate condensed matter, nuclear, and particle physics. In the case of gauge theories comprising the Standard Model (SM) of particle physics, progress in perturbative tools has enabled predictions for high-energy experiments at the Large Hadron Collider~\cite{Tanabashi:2018oca}. Furthermore, progress in non-perturbative tools has led to theoretical input for precision experiments in search of violations of fundamental symmetries in nature, and to predicting hadronic excitations and their internal structure~\cite{Aoki:2019cca}. Nonetheless, the computational complexity of such studies grows significantly with the system size. In the strong-coupling regime, in which non-perturbative Monte Carlo sampling of quantum vacuum configurations is a common practice, questions such as the nature of the phase diagram of finite-density systems and the real-time dynamics of matter remain unanswered. It is therefore essential to explore a broader set of computational approaches, including those based on quantum simulation and quantum computation~\cite{nielsen2000quantum,lloyd1996universal,georgescu2014quantum}, to tackle these problems.

While the idea of simulating a quantum system using another quantum system with a higher level of control dates back to Feynman~\cite{Feynman:1982}, only the experimental advancements in recent years have enabled powerful and sizable quantum simulations to become a reality. As in the case of classical computations, digital computations on quantum platforms may be the ultimate solution to all computational problems, including quantum simulations of physical systems. However, in the era of noisy intermediate-scale quantum (NISQ) computing~\cite{preskill2018quantum}, the number of high-fidelity operations that can be performed on a device can be highly constrained by the short coherence time of the quantum state. As a result, the digitalization of complex dynamics, such as those associated with gauge field theories, can be limited to small system sizes and short evolution times. It is therefore important to seek alternative approaches in the NISQ era. An interesting possibility is offered by analog simulations, in which the native Hamiltonian of the controlled quantum system is engineered to be mapped to that of the target system. The quantum operations are then naturally implemented once the system is prepared to evolve according to the desired Hamiltonian.

Among the most compelling platforms for analog simulations of quantum systems, including those governed by gauge theories, are cold neutral atoms in optical lattices~\cite{jaksch1998cold,greiner2002quantum,weimer2010rydberg,lewenstein2012ultracold,bloch2012quantum,cirac2012goals,gross2017quantum,arguello2018analog}, optical tweezers~\cite{bernien2017probing,de2018experimental}, and trapped ions~\cite{blatt2012quantum,wineland1998experimental}. Simple low-dimensional field theories such as relativistic Dirac fermions, 1+1D\footnote{Here and in the following, the first number denotes the space dimension, and the second number refers to the time dimension. When there is only one number, it is meant to refer to the space dimension (or the spacetime dimension with a Euclidean metric).} and 2+1D scalar and fermionic quantum electrodynamics (QED), and non-Abelian SU(2) and SO(3) gauge theories have been studied in this context, and proposals exist to map the desired lattice Hamiltonians (or their approximated forms) to that of the engineered Hamiltonian of neutral atoms in optical lattices~\cite{cirac2010cold,Boada:2010sh,buchler2005atomic,Zohar:2011cw,szirmai2011gauge,Zou:2014rha,Tagliacozzo:2012df,Zohar:2012xf,Banerjee:2012pg,Banerjee:2012xg,Zohar:2012ts,Zohar:2013zla,Wiese:2013uua,Rico:2013qya,Wiese:2014rla,Zohar:2015hwa,Dalmonte:2016alw,Notarnicola:2015sia,Kasper:2016mzj,Kuno:2016ipi,Rico:2018pas,Zhang:2018ufj}. Recent implementations of simple static and dynamical gauge theories with neutral atoms in optical lattices~\cite{dalibard2011colloquium,mancini2015observation,stuhl2015visualizing,Aidelsburger:2017qlh,clark2018observation,schweizer2019floquet}, however, demonstrate the challenge of simulating more phenomenologically-relevant gauge theories. Given the current size of controlled quantum systems, only a small number of degrees of freedom can be studied, leading to unavoidable truncations in the Hilbert space of a gauge theory that lives in a continuous infinite-volume spacetime. Such a limitation is present in other digital and analog quantum platforms as well. It is nonetheless important that theoretical developments in formulating and mapping gauge theories for a quantum simulation proceed alongside the continual experimental progress that aims to significantly improve capabilities and capacities of simulating platforms.

Trapped ions provide a pristine platform for quantum simulations~\cite{blatt2012quantum}. Given the extremely high level of control enabled by laser-cooled and localized ions confined by electromagnetic fields, exceedingly high fidelities in state preparation and measurement, all-to-all entangling capability enabled through control over the excitations of the motional normal modes, and scalability potential of such systems, this architecture has become a primary candidate for digital quantum computations in recent years~\cite{cirac1995quantum,molmer1999multiparticle,solano1999deterministic,milburn2000ion,blatt2008entangled,lanyon2011universal,monroe2013scaling,monroe2014large,Martinez:2016yna,debnath2016demonstration,linke2017experimental,figgatt2018parallel,landsman2019verified,nam2019ground,wright2019benchmarking}. A unique feature of the trapped ion architecture is that global addressing of the ions using a few laser beams allows the realization of tunable long-range spin-spin interactions in the chain. With no need for individual addressability, systems of a few tens of ions have been successfully realized, and analog simulations of sizable quantum spin systems are made possible~\cite{porras2004effective,islam2011onset,schneider2012experimental,jurcevic2014quasiparticle,richerme2014non,zhang2017observation,hess2017non,neyenhuis2017observation,liu2018confined}. More complex quantum many-body systems, such as those described by gauge field theories, require either some degree of individual addressing or higher-order spin interactions among different species, as put forward in several proposals for simulating the relativistic Dirac equation~\cite{lamata2007dirac,gerritsma2010quantum,casanova2010klein,georgescu2014quantum} a quantum field theory of scalar fields~\cite{Casanova:2011wh}, and 1+1D QED~\cite{Hauke:2013jga,Martinez:2016yna}. A milestone in quantum simulations of lattice gauge theories (LGTs) using trapped ions was achieved in Ref.~\cite{Martinez:2016yna}, where the real-time dynamics of 1+1D QED in a system of four trapped ions in a linear trap was made possible through a digital protocol, but the number of operations required for a Trotterized procedure prevented a long evolution time to be achieved in the presence of noise. While fully analog proposals exist for simulating simple low-dimensional LGTs~\cite{Hauke:2013jga}, none have been implemented so far due to technical limitations.

It is important to classify gauge field theories of interest in terms of whether analog simulation of their dynamics is feasible given current technology. It is also essential to investigate whether fully analog implementations can circumvent the accumulated noise due to digitalization~\cite{suzuki1976generalized}, and whether the noise in an analog setup can be effectively mitigated. Finally, it would be beneficial to assess the practicality of existing ideas, and to develop new proposals for extending the quantum toolkit of trapped ions, to enable a one-to-one mapping between the engineered Hamiltonian of the ion-laser quantum system to the dynamics of a fermionic system coupled to gauge degrees of freedom (bosons). This paper is a first step in addressing these questions. Here, we focus on identifying goals that can be achieved in the near term, by specifying, in detail, practical proposals for a range of accessible gauge theories.

The gauge field theories studied in this paper are examples of the theories whose discretized formulations can be mapped entirely to systems with spin-$\frac{1}{2}$ degrees of freedom. These examples include: i) 1+1D quantum electrodynamics (Schwinger model). This model has similarities to quantum chromodynamics in 3+1D, including exhibiting a non-trivial vacuum. ii) 2+1D Abelian Chern-Simons theory coupled to matter fields. This model is an example of a topological gauge theory with applications in many areas of physics. iii) 2+1D $Z_2$ gauge theory with a non-trivial phase diagram on a lattice, including exhibiting confinement. We discuss the mapping of these theories to spin systems, and present experimental protocols for realizing these interactions in current and near-term ion-trap systems. In order to provide a reference for upcoming implementations in the case of the Schwinger model, detailed examples for 4 and 8 fermion-site theories will be presented.

A linear chain of trapped ions is often viewed as a platform for simulating spin-$\frac{1}{2}$ systems in 1+1D. However, once such a system is augmented with individual addressing, it offers far more possibilities for quantum simulations of arbitrary spin systems, including those in higher dimensions. Such proposals have been put forth in Ref.~\cite{korenblit2012quantum}, and are explicitly taken advantage of in the current work to simulate the dynamics of the LGTs mentioned above. We also demonstrate the accessibility of nearly-perfect nearest-neighbor interactions by simply controlling the lasers' phases and intensities on each ion, and demonstrate the sensitivity of the evolution to the imperfections of the engineered Hamiltonian in the case of the Schwinger model. By controlling intensities, phases and frequencies of laser beams addressing each ion, a highly accurate mapping to spin-spin Hamiltonians with arbitrary interaction profiles is enabled. An important feature of the protocols devised in this work is a thorough optimization procedure that maximizes the closeness to the desired Hamiltonian, while simultaneously minimizes errors stemming from residual couplings to motional excitations. The proposed experimental scheme will have applications beyond the examples discussed, and is a general protocol for realizing interesting spin systems described by a Heisenberg Hamiltonian in arbitrary dimensions. 

The paper is organized as follows. Sec.~\ref{sec:ionphysics} includes details of ion-laser Hamiltonian considered in the scheme of this work, and presents the effective Hamiltonian obtained, its range of validity, and the associated undesired contributions that must be minimized subsequently. The two associated appendices~\ref{app:Yb} and \ref{app:Rabi} offer details on a particular experimental platform, and a scheme that eliminates an unwanted bias term in engineering the effective Hamiltonian. The full evolution operator is further detailed in Appendix~\ref{app:H}. Sec.~\ref{sec:schwinger} presents the example of the lattice Schwinger model, its purely spin representation, and explicit experimental proposals for simulating 4 and 8 fermion-site theories. The former case is implemented with a single detuning for each set of the lasers used, while the latter takes advantage of a multi-frequency, multi-amplitude scheme, requiring a thorough optimization of interaction couplings. Additional results on the 8 fermion-site theory are presented in Appendix~\ref{app:N8plots}. The results of the numerical evaluation of the full evolution operator up to the order considered are presented in another associated Appendix (Appendix \ref{app:numerics}) as well as in Supplemental Material. Sec.~\ref{sec:higherd} presents examples of LGTs in higher dimensions and their dual spin representation, along with discussions on their amenability to the quantum simulation scheme of this work. We conclude in Sec.~\ref{sec:conclusion} by highlighting the differing  features of the scheme presented here compared with the previous work, the significance of the results obtained, and future extensions that may enable addressing a wider class of gauge theories.

\section{1D chain of trapped ions and engineered effective interactions
\label{sec:ionphysics}}
\noindent
Consider $N$ ions confined in a radio-frequency Paul trap~\cite{paul1990electromagnetic}. The ``qubit'' in this system can be encoded in two stable internal levels of the ion, denoted in the following as $\ket{\uparrow}$ and $\ket{\downarrow}$. These states are separated in energy by an angular frequency $\omega_0$ (with Planck's constant $\hbar = 1$ here and in the rest of the paper). Coherent operations on spin degrees of freedom are realized through stimulated Raman transitions using two laser beams with a momentum-vector difference $\Delta \bm{k}$. The physics of ion-laser interactions and the single and two-qubit manipulations in an ion trap is well known~\cite{wineland1998experimental,porras2004effective,zhu2006trapped,zhu2006arbitrary,schneider2012experimental,doi:10.1142/p983}. However, the involved evolution of the system under multiple pairs of Raman beams, which are needed for engineering the Hamiltonians of models considered here, requires a few technical novelties, and warrants a dedicated discussion which will follow in this section. For presentational clarity, further details of the proposed scheme and a number of involved analytical forms will be presented in the appendices.

The ion-laser interaction Hamiltonian for a system of $N$ trapped ions can be written as~\cite{schneider2012experimental}  
\begin{eqnarray}
&&H_{\rm int}=\sum_{i=1}^N \sum_{L=1}^{n_L} \Omega_L^{(i)} e^{-i\Delta\omega_L^{(i)}t+i\Delta\varphi_L^{(i)}+i\Delta \bm{k}_L^{(i)} \cdot \Delta \bm{r}^{(i)}}
\nonumber\\
&&  \hspace{0.35 in} \times ~ (\alpha_0\mathbb{I}^{(i)}+\alpha_1\sigma_x^{(i)}+\alpha_2\sigma_y^{(i)}+\alpha_3\sigma_z^{(i)})+\text{h.c.},
\label{eq:Hionlaser}
\end{eqnarray}
Index $L$ in Eq.~(\ref{eq:Hionlaser}) runs over $n_L$ pairs of Raman beams. $\Omega^{(i)}_L$ is the Rabi frequency associated with the laser $L$. $\Delta\varphi_L^{(i)}$ denotes the phase difference between the two lasers in each pair of Raman beams, $\Delta\omega_L^{(i)}$ is the difference in their angular frequency, namely the beatnote frequency, and $\Delta \bm{k}_L^{(i)}$ is the difference in their momentum $k$-vector. In general, each ion is addressed with multiple pairs of Raman beams individually (hence the superscript $(i)$ on quantities), requiring both amplitude and frequency control of the beams. Such individual addressing of the ions is widely used in digital ion-trap platforms, and can be ported to analog platforms in upcoming experiments. $\Delta \bm{r}^{(i)}$ denotes the displacement vector of ion $i$ from its equilibrium position. The Pauli matrices $\sigma^{(i)}$ act on the quasi-spin of ion $i$, and $\alpha_0,\alpha_1,\alpha_2$, and $\alpha_3$ are constants related to the spin-dependent forces on the two states of the qubit~\cite{schneider2012experimental} and are controlled by the intensity, geometry and polarization of the laser beams, see Appendix~\ref{app:Rabi} for further details.

\begin{figure*}[t!]
\includegraphics[scale=0.45]{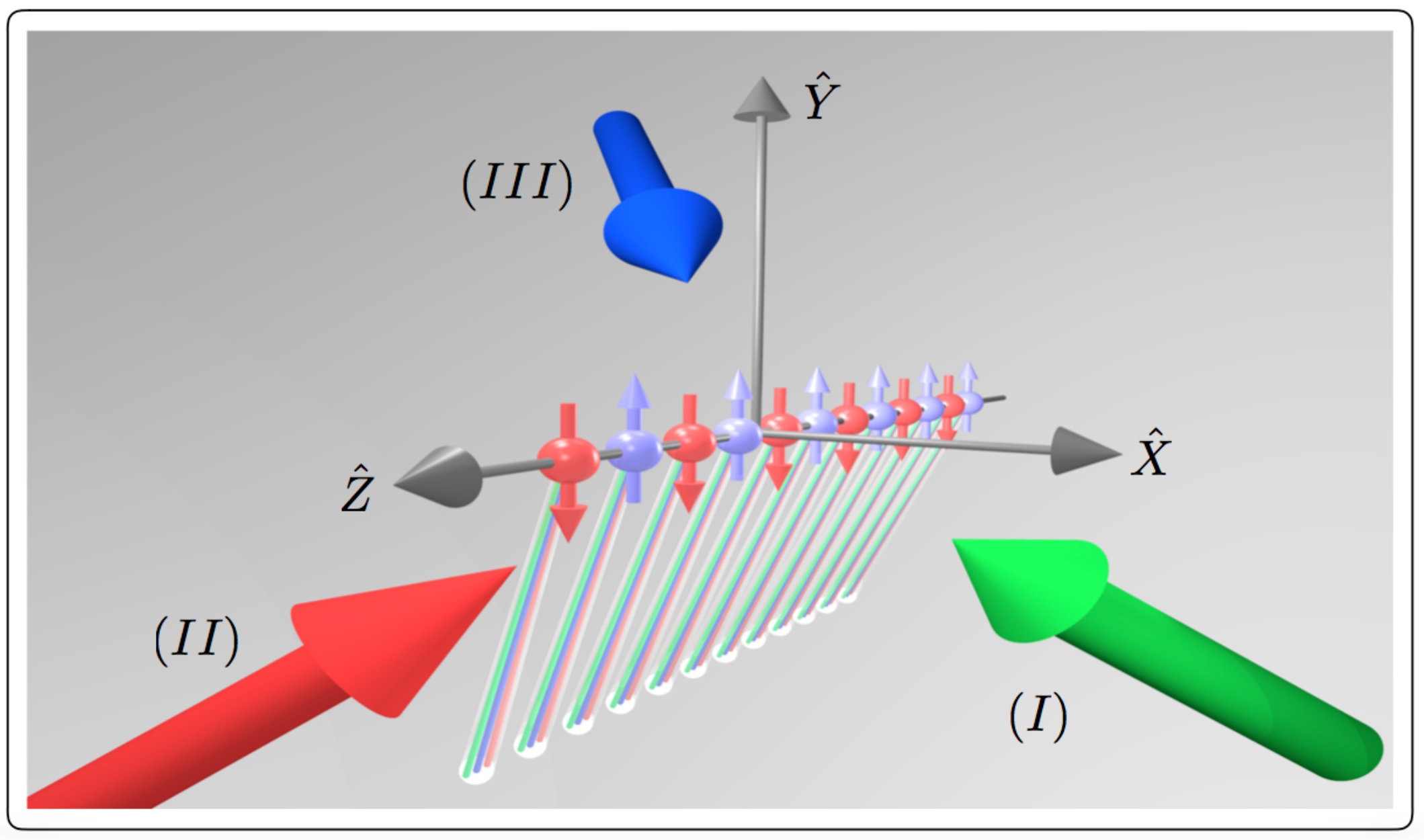}
\caption[.]{A schematic representation of a Raman-beams configuration that induces effective spin-spin interactions in the Heisenberg model. The $N$ sets of individual beams can be chosen along the $(\xi\hat{X},\xi\hat{Y},\chi\hat{Z})$ unit vector ($2\xi^2+\chi^2=1$). Global beams $(I)$, $(II)$, and $(III)$ are then chosen to propagate along $(-\xi\hat{X},\xi\hat{Y},\chi\hat{Z})$, $(\xi\hat{X},\xi\hat{Y},-\chi\hat{Z})$, and $(\xi\hat{X},-\xi\hat{Y},\chi\hat{Z})$, respectively. These will cause net $\Delta \bm{k}$ vectors compared with the individual beams along the $\hat{X}$, $\hat{Z}$, and $\hat{Y}$ directions, respectively. Chosen values of these parameters for the examples of this work are given in Appendix~\ref{app:Yb}.
}
\label{fig:lasers}
\end{figure*}

We assume that the confining potential is sufficiently stronger along the transverse axes of the trap so that the ions form a 1D crystal in space.  With appropriate anharmonic axial confinement forces, the ions can be nearly equally-spaced~\cite{lin2009large,pagano2018cryogenic}, with a typical spacing between adjacent ions of a few micrometers. Due to the long-range Coulomb force among the ions and the common trapping potential applied, the motion of the ions can be described in terms of a set collective normal modes. Then, $\Delta \bm{r}^{(i)}$ in Eq.~(\ref{eq:Hionlaser}) can be expressed in terms of phononic degrees of freedom, whose excitation energies are quantized in units of the normal-mode frequencies of the system. For the Hamiltonians of gauge theories considered in this work, it is necessary to introduce multiple pairs of bichromatic Raman beams directed at each ion, such that each pair couples to only one set of the three independent sets of normal modes. Such a scheme can be achieved with $N$ individual beams and three global beams. Each of the individual beams will have three frequencies\footnote{Or three sets of frequencies as required by the multi-frequency scheme of Sec.~\ref{sec:schwinger}.} that are tuned sufficiently apart such that each frequency will drive the qubit only by pairing with one of the global beams. This setup will allow to tune independently Hamiltonians acting along orthogonal directions of the Bloch sphere with negligible undesired cross couplings as shown below. The chosen directionality of the beams can ensure that each global-individual pair will result in a net $k$-vector along one of the three orthogonal principal axes of the trap, $X,Y$ and $Z$, see Fig.~\ref{fig:lasers}.\footnote{These Cartesian indices must not be confused with the $x,y$ and $z$ indices introduced on quasi-spins of the qubit. While the former (upper-case letters) correspond to the components of laser fields' $k$-vector, the latter (lower-case letters) correspond to the Bloch-sphere axes in the qubit Hilbert space} Here, $X$ and $Y$ denote the most-confined directions in the trap, which will have the same normal-mode spectra for symmetric traps commonly used. These will be denoted as transverse directions. The least-confined direction is denoted as $Z$ and is named the axial direction.

Consider now the ion-laser system in the interaction picture, in which all excitations arising from the free Hamiltonian
\begin{eqnarray}
&& H_0 = 
\sum_{i=1}^N \frac{\omega_0}{2} \sigma_z^{(i)} + \sum_{m=1}^N \left[ \omega_m^T (a_m^\dagger a_m+\frac{1}{2})+ \right .
\nonumber\\
&& \hspace{0.3 cm} \left . \omega_m^A (b_m^\dagger b_m+\frac{1}{2})+ \omega_m^T (c_m^\dagger c_m+\frac{1}{2})\right] + \text{const.}
\label{eq:H0}
\end{eqnarray}
are rotated away by frequencies of the order of $\omega_0$, $\omega_m^T$, and $\omega_m^A$.\footnote{Although the axial modes are generally low in frequency, such a rotating-frame approximation is still valid as long as lasers' detunings from these modes remain small compared to the sideband Rabi frequencies of the axial motion.} $a_m$ ($a_m^{\dagger}$) annihilates (creates) a phonon excitation of the transverse normal mode $m$ with angular frequency $\omega_m^T$ along the $X$ direction of the trap, i.e., $\Delta \bm{k}_I=\Delta k_I \hat{\bm{X}}$. Similarly, $b_m$ and $c_m$ ($b_m^{\dagger}$ and $c_m^{\dagger}$) are, respectively, the phonon annihilation (creation) operators for the axial normal modes along the $Z$ direction, i.e., $\Delta \bm{k}_{II}=\Delta k_{II} \hat{\bm{Z}}$ and the transverse normal modes along the $Y$ direction, i.e., $\Delta \bm{k}_{III}=\Delta k_{III} \hat{\bm{Y}}$.\footnote{At this point, such assignments of a given set of normal modes to one of the Hamiltonians in Eqs.~(\ref{eq:HintI}-\ref{eq:HintIII}) appear arbitrary. The rationale behind the choices made will become clear in applications of the scheme to nearest-neighbor Hamiltonians considered in this work, see Sec.~\ref{sec:schwinger}.} The corresponding normal mode frequencies are denoted as $\omega_m^A$ and $\omega_m^T$. Different superscripts are introduced to distinguish the transverse and axial normal modes which have different spectra. Finally, in the Lamb-Dicke regime where $\langle\Delta k^{(i)} \Delta r^{(i)}\rangle \ll 1$, and when the laser frequencies are chosen such that all transitions except for those near the first sideband transitions\footnote{The $n^{\text{th}}$ blue (red) sideband transition for mode $m$ adds (removes) $n$ quanta of motion each with frequency $\omega_m$.} are far off-resonant, the three sets of Raman-beam pairs at each ion induce the laser-ion Hamiltonians of the form
\begin{eqnarray}
\label{eq:HintI}
&&\widetilde{H}_I=\sum_{i=1}^N i\Omega_I^{(i)}\left( e^{i \mu_It+i\Delta\varphi_I^{(i)}}+e^{-i \mu_It+i\Delta\varphi_I^{\prime(i)}}\right)\times\sum_{m=1}^N
\nonumber\\
&& \hspace{0.25 cm} \eta_{I,m}^{(i)}  \left(a_me^{-i\omega_m^Tt}+a_m^\dagger e^{i\omega_m^Tt} \right)(\alpha_1-i\alpha_2)\sigma_+^{(i)}+\text{h.c.},
\nonumber\\
\end{eqnarray}
\begin{eqnarray}
\label{eq:HintII}
&&\widetilde{H}_{II}=\sum_{i=1}^N i\Omega_{II}^{(i)}\left( e^{i \mu_{II}t+i\Delta\varphi_{II}^{(i)}}+e^{-i \mu_{II}t+i\Delta\varphi_{II}^{\prime(i)}}\right) \sum_{m=1}^N
\nonumber\\
&& \hspace{0.25 cm} \eta_{II,m}^{(i)} \left(b_me^{-i\omega_m^At}+b_m^\dagger e^{i\omega_m^At} \right)(\alpha_1-i\alpha_2)\sigma_+^{(i)}+\text{h.c.},
\nonumber\\
\end{eqnarray}
\begin{eqnarray}
&&\widetilde{H}_{III}=\sum_{i=1}^N i\Omega_{III}^{(i)}\left( e^{i \mu_{III}t+i\Delta\varphi_{III}^{(i)}}+e^{-i \mu_{III}t+i\Delta\varphi_{III}^{\prime(i)}}\right)
\nonumber\\
&& \hspace{0.0 cm} \sum_{m=1}^N \eta_{III,m}^{(i)} \left(c_me^{-i\omega_m^Tt}+c_m^\dagger e^{i\omega_m^Tt} \right)(\alpha_0 \mathbb{I}^{(i)}+\alpha_3\sigma_z^{(i)})+\text{h.c.},
\nonumber\\
\label{eq:HintIII}
\end{eqnarray}
where $\sigma_{\pm}^{(i)}=\frac{1}{2}(\sigma_x^{(i)} \pm i\sigma_y^{(i)})$, and tilde over the Hamiltonians imply the use of the rotated frame described above. Here, it is assumed that $|\mu_I| \ll \omega_0$ where $\mu_I \equiv \omega_0-\Delta\omega_I=-\omega_0+\Delta\omega^\prime_I$. Similarly, $|\mu_{II}| \ll \omega_0$ where $\mu_{II} \equiv \omega_0-\Delta\omega_{II}=-\omega_0+\Delta\omega_{II}^\prime$. On the other hand, for the Hamiltonian $\widetilde{H}_{III}$, it is assumed that $|\mu_{III}| \ll \omega_0$ where $\mu_{III} \equiv -\Delta\omega_{III}=\Delta\omega_{III}^\prime$. Further, two distinct Raman-beam phase differences are assigned to each of the red (unprimed) and blue (primed) detuned frequencies of the beam. $\eta_{I,m}^{(i)}=\sqrt{\frac{(\Delta k_I)^2}{2M\omega_m^T}}b_m^{(i)}$, where $b_m^{(i)}$ is the (normalized) normal-mode eigenvector components between ion $i$ and mode $m$, and $M$ denotes the mass of the ion. Similarly, $\eta_{II,m}^{(i)}=\sqrt{\frac{(\Delta k_{II})^2}{2M\omega_m^A}}b_m^{(i)}$ and $\eta_{III,m}^{(i)}=\sqrt{\frac{(\Delta k_{III})^2}{2M\omega_m^T}}b_m^{(i)}$ for the axial and transverse modes, respectively. For each pair of Raman beams $L$, the same $\Delta k_L$-vector is applied at the location of each ion. $\alpha_1=\frac{1}{2}$ and $\alpha_2=0$ correspond to the well-known Molmer-Sorenson scheme, already applied in a number of experiments. In order to eliminate a bias $\sigma_z$ interaction arising from $\widetilde{H}_{III}$, it is essential that $\alpha_0$ is set to zero. With the scheme presented in Appendix~\ref{app:Rabi}, it is shown that one can achieve this requirement by tuning the Raman-beam frequencies and polarization vectors. We further set $\alpha_3=\frac{1}{4}$ for consistency between the effective spin-spin couplings arising from $\widetilde{H}_I$, $\widetilde{H}_{II}$, and $\widetilde{H}_{III}$.\footnote{There will be no ambiguity in the overall constants in the Hamiltonian. Rescaling these coefficients by a constant means the Rabi frequencies must be rescaled accordingly so that the expected strength of the state-dependent force is produced on a given ion, and with given choices of the internal levels for the qubit.} Now by setting the phases of the blue and red-sideband detuned beams to $\Delta\varphi_I^{(i)}=0$, $\Delta\varphi_I^{\prime(i)}=\pi$, $\Delta\varphi_{II}^{(i)}=\Delta\varphi_{II}^{\prime(i)}=\Delta\varphi_{III}^{(i)}=\Delta\varphi_{III}^{\prime(i)}=0$, the Hamiltonians $\widetilde{H}_I$, $\widetilde{H}_{II}$, and $\widetilde{H}_{III}$ can be seen to be proportional to $\sigma_x^{(i)}$, $\sigma_y^{(i)}$, and $\sigma_z^{(i)}$, respectively.

Finally, an effective longitudinal magnetic field can be introduced at the location of each ion by another $N$ sets of beams inducing a Stark shift to be tuned to the desired value of the magnetic field. Alternatively, a $B_z$ field can be generated with the existing sets of Raman beams, i.e., by shifting the frequency of red and blue-detuned beams by $B_z^{(i)}$. This can be seen by noting that if the rotating frame that led to Eqs.~(\ref{eq:HintI}-\ref{eq:HintIII}) is assumed to rotate with the Hamiltonian $H_0+\frac{1}{2}\sum_{i=1}^N B_z^{(i)} \sigma_z^{(i)}$ instead of $H_0$, in addition to the interacting Hamiltonians in Eqs.~(\ref{eq:HintI}-\ref{eq:HintIII}), an effective Hamiltonian
\begin{eqnarray}
H_B&=&-\frac{1}{2}\sum_{i=1}^N B_z^{(i)} \sigma_z^{(i)}
\label{eq:HB}
\end{eqnarray}
is generated, but at the cost of the following change: $\mu_I \to \mu_I + B_z^{(i)}$ and $\mu_I \to \mu_I - B_z^{(i)}$ to the laser detuning in the first and second occurrences of $\mu_I$ in Eq.~(\ref{eq:HintI}), respectively. Similarly, $\mu_{II}$ must be replaced by $\mu_{II} \to \mu_{II} + B_z^{(i)}$ and $\mu_{II} \to \mu_{II} - B_z^{(i)}$ in the first and second occurrences in Eq.~(\ref{eq:HintII}), respectively. The laser detuning $\mu_{III}$, on the other hand, remains unchanged. Note that this scheme requires a frequency control, as the detunings are now generally different at the location of each ion.

With the Hamiltonians in Eqs.~(\ref{eq:HintI}-\ref{eq:HB}), an evolution operator can be formed by applying a Magnus expansion, taking into account all contributions up to and including $\mathcal{O}\left(\eta^2,\eta B_z\right)$ in the exponent:
\begin{eqnarray}
U(t,0)&=&\exp\left[\sum_{\alpha=x,y,z}\left(\sum_{i=1}^N ~ \phi_i^{(\alpha)}(t)~\sigma_\alpha^{(i)} \ + \ \right . \right .
\nonumber\\
&& \hspace{0.65 in} \left . \left . \sum_{i,j} ~ \chi^{(\alpha)}_{i,j}(t)~\sigma_\alpha^{(i)}\otimes \sigma_\alpha^{(j)}\right) \right],
\label{eq:UAll}
\end{eqnarray}
where
\begin{eqnarray}
\label{eq:phiX}
\phi_i^{(x)}(t) &=& \sum_{m=1}^N \alpha^{(x)}_{i,m}(t)~a_m^{\dagger}+\sum_{m=1}^N\sum_{n=1}^N  \beta^{(x)}_{i,m,n}(t)~b_m^{\dagger}c_n+\text{h.c.},
\nonumber\\
\\
\label{eq:phiY}
\phi_i^{(y)}(t) &=& \sum_{m=1}^N \alpha^{(y)}_{i,m}(t)~b_m^{\dagger}+\sum_{m=1}^N\sum_{n=1}^N  \beta^{(y)}_{i,m,n}(t)~a_m^{\dagger}c_n+\text{h.c.},
\nonumber\\
\\
\label{eq:phiZ}
\phi_i^{(z)}(t) &=& 
 \gamma^{(z)}_{i}(t)+\sum_{m=1}^N \alpha^{(z)}_{i,m}(t)~c_m^{\dagger}+\hspace{1.5 cm}
\nonumber\\
&&\hspace{1.75 cm} \sum_{m=1}^N\sum_{n=1}^N \beta^{(z)}_{i,m,n}(t)~b_m^{\dagger}a_n+\text{h.c.}
\end{eqnarray}
The definition of the rest of the functions in Eqs.~(\ref{eq:UAll}-\ref{eq:phiZ}) are provided in Appendix~\ref{app:H}.

When $B_z^{(i)}=0$, all contributions proportional to phonon creation and annihilation operators in the exponent in Eq.~(\ref{eq:UAll}) are bounded in time, provided that $\mu_I \neq \mu_{II} \neq \mu_{III}$. As a result, an effective Heisenberg model can be achieved when $t \gg {|\mu_I - \omega_m^T|}^{-1},~{|\mu_{II} - \omega_m^A|}^{-1},~{|\mu_{III} - \omega_m^T|}^{-1}$, so that the terms linear in time in Eq.~(\ref{eq:UAll}) (those proportional to $\chi_{i,j}^{(\alpha)}$) dominate the evolution. In such a limit, $\chi_{i,j}^{(\alpha)} \to -\frac{i}{2}J_{i,j}^{(\alpha\alpha)}t$ (see Eqs.~(\ref{eq:JXXdef}-\ref{eq:JZZdef})), and other contributions will be subdominant. For practical (noisy) implementations, one needs to minimize the spin-phonon entanglement arising from the first term in the exponent in Eq.~(\ref{eq:UAll}) at early times. This is achieved with $|\eta_{I,m}^{(i)} \Omega_I^{(i)}| \ll |\mu_I - \omega_m^T|$, $|\eta_{II,m}^{(i)} \Omega_{II}^{(i)}| \ll |\mu_{II} - \omega_m^A|$, and $|\eta_{III,m}^{(i)} \Omega_{III}^{(i)}| \ll |\mu_{III} - \omega_m^T|$. When $B_z^{(i)} \neq 0$, $\alpha^{(x)}_{i,m}(t)$, and $\alpha^{(y)}_{i,m}(t)$ in Eqs.~(\ref{eq:phiX}) and (\ref{eq:phiY}) develop an oscillatory time dependence but with a linear growth in the magnitude of its amplitude. These terms are proportional to $B_z^{(i)}\sigma_y^{(i)}$ and $B_z^{(i)}\sigma_x^{(i)}$. Assuming that the magnetic field is comparable in size to the effective spin-spin couplings, such contaminating terms do not severely impact the desired evolution as long as $|B_z^{(i)}| \ll |\eta_{I,m}^{(i)} \Omega_I^{(i)}|,~|\eta_{II,m}^{(i)} \Omega_{II}^{(i)}|$. Unfortunately, this condition limits the size of (effective) magnetic fields that can be studied in models considered below. Nonetheless, a range of interesting possibilities can still be explored.

Under the conditions described above, the time-evolution operator in Eq.~(\ref{eq:UAll}) can be approximated as
\begin{eqnarray}
U(t) \approx e^{-i H_{\text{eff}}t},
\label{eq:Ueff}
\end{eqnarray}
where
\begin{eqnarray}
H_{\text{eff}} 
&=& \sum_{\underset{j<i}{i,j}} \left[J_{i,j}^{(xx)}\sigma_x^{(i)} \otimes \sigma_x^{(j)}+J_{i,j}^{(yy)}\sigma_y^{(i)} \otimes \sigma_y^{(j)}+\right .
\nonumber\\
&& \hspace{0.6 cm} \left . J_{i,j}^{(zz)}\sigma_z^{(i)} \otimes \sigma_z^{(j)} \right] - \frac{1}{2}\sum_{i=1}^N B_z^{(i)}\sigma_z^{(i)}.
\label{eq:Heisenberg}
\end{eqnarray}
As a result, the individual-addressing scheme proposed here enables analog quantum simulations of a rather generic Heisenberg spin model. The spin-spin coupling matrices in Eq.~(\ref{eq:Heisenberg}) are derived from discussions above (see also Appendix~\ref{app:H}), and read
\begin{eqnarray}
J_{i,j}^{(xx)} 
&=& \Omega_{I}^{(i)}\Omega_{I}^{(j)} R_I \sum_{m=1}^N\frac{b_m^{(i)}b_m^{(j)}}{\mu_I^2-{\omega_m^T}^2},
\label{eq:JXXdef}
\\
J_{i,j}^{(yy)} &=& \Omega_{II}^{(i)}\Omega_{II}^{(j)} R_{II} \sum_{m=1}^N\frac{b_m^{(i)}b_m^{(j)}}{\mu_{II}^2-{\omega_m^A}^2},
\label{eq:JYYdef}
\\
J_{i,j}^{(zz)} &=& \Omega_{III}^{(i)}\Omega_{III}^{(j)} R_{III} \sum_{m=1}^N\frac{b_m^{(i)}b_m^{(j)}}{\mu_{III}^2-{\omega_m^T}^2}.
\label{eq:JZZdef}
\end{eqnarray}
Here, $R_L = \frac{(\Delta k_L)^2}{2M}$ is the recoil frequency of the ion given the lasers $L=I,II,III$.

It is worth noting that despite the case of a usual Molmer-Sorenson transition where the starting Hamiltonian is proportional to $\sigma_x$, the Magnus expansion in the scheme described above is not cut off at any order in the Lamb-Dicke parameter, due to the non-zero commutation of Pauli operators in Eqs.~(\ref{eq:HintI}-\ref{eq:HB}). It is therefore important to ensure that not only $\left|\frac{\eta_{I,m}^{(i)} \Omega_I^{(i)}}{\mu_I - \omega_m^T}\right|,~\left|\frac{\eta_{II,m}^{(i)} \Omega_{II}^{(i)}}{\mu_{II} - \omega_n^A}\right|,~\left|\frac{\eta_{III,m}^{(i)} \Omega_{III}^{(i)}}{\mu_{III} - \omega_m^T}\right| \ll 1$ as stated before, but also $\left|\frac{(\eta_{I,m}^{(i)})^{(2p-2)}(\mu_I - \omega_m^T)}{\mu_I - p\omega_m^T}\right|,~\left|\frac{(\eta_{II,m}^{(i)})^{(2p-2)}(\mu_{II} - \omega_m^A)}{\mu_{II} - p\omega_m^A}\right|,$ $\left|\frac{(\eta_{III,m}^{(i)})^{(2p-2)}(\mu_{III} - \omega_m^T)}{\mu_{III} - p\omega_m^T}\right| \ll 1$ for integer $p \geq 2$. This guarantees that contributions from the $p^\text{th}$-sideband transitions are suppressed compared to the first-sideband transitions. These conditions are easier to satisfy for transverse modes than the axial modes. This is because the axial modes have lower frequencies, and their corresponding Lamb-Dicke parameters are larger. Finally, one notes that coherent operations on a single spin correspond to the zeroth-order terms in Eq.~(\ref{eq:Hionlaser}) in the Lamb-Dicke limit, and with $\Delta \omega_L^{(i)} = \omega_0$. Hence, the laser frequencies applied must be far detuned from such ``carrier transitions'' of the ions.

\section{Optimized spin-spin Hamiltonians in an ion trap: 1+1D Schwinger Model
\label{sec:schwinger}}
\noindent
A unique testbed for exploring theoretical and experimental proposals for quantum simulations of gauge theories is the $1+1$D QED, i.e., the Schwinger model. It is an Abelian gauge theory, hence avoiding complexities of its non-Abelian counterparts. It is also a low-dimensional theory, allowing numerical and experimental studies of its approximate dynamics with finite resources. Despite these simplifications in the formulation, the theory exhibits rich properties, similar to those seen in more complex theories such as QCD. In particular, phenomena such as confinement and spontaneous symmetry breaking arise in the model. The spontaneous creation of electron-positron pairs in the time evolution of the ``vacuum'' exhibits a clear signature of such non-trivial dynamics. Since the time evolution of quantum states is, in general, a computationally intractable problem with classical Monte Carlo methods, addressing such a problem using a quantum simulation platform is of significant value. 

The strong-coupling dynamics of the Schwinger model can be studied through non-perturbative LGT methods. In the staggered formulation of Kogut and Susskind~\cite{kogut1975hamiltonian, banks1976strong}, the (scaled) lattice Hamiltonian takes the form
\begin{eqnarray}
&&H = -ix \sum_{n=1}^{N-1} \left[ \Phi_n^{\dagger} e^{i\theta_n }\Phi_{n+1}-\Phi_{n+1}^{\dagger} e^{-i\theta_n }\Phi_n \right]+
\nonumber\\
&&\hspace{1 in} \sum_{n=1}^{N-1} L_n^2+\mu \sum_{n=1}^N (-1)^n \Phi_n^{\dagger} \Phi_n,
\label{eq:HSchwingerI}
\end{eqnarray}
where $\Phi_n$ ($\Phi_n^{\dagger}$) is a one-component fermion field that creates (annihilates) an electron on the odd site while annihilates (creates) a positron on an even site. Due to this distinction,  there is a staggered mass term in the Hamiltonian, with the fermion (scaled) mass $\mu$. $\theta_n$ is the $U(1)$ gauge potential with the corresponding gauge link $e^{i\theta_n}$ originating at site $n$. The latter is introduced in the Hamiltonian to render the fermion hopping (kinetic) term gauge invariant. The pair creation and annihilation in the theory originates from this term. The corresponding electric field at site $n$ is denoted as $L_n$ (with the operator relation $[\theta_n,L_m]=i\delta_{n,m}$), which adds a contribution to the Hamiltonian due to the energy stored in the electric field. The Hamiltonian in Eq.~(\ref{eq:HSchwingerI}) is written in units of $ag^2/2$, where $a$ denote the lattice spacing and $g$ is the original fermion-gauge field coupling. The dimensionless parameters $x$ and $\mu$ are related to dimensionful parameter $g$ (with mass dimension one) and the original mass $m$ via: $x=1/(ag)^2$ and $\mu=2m/(ag^2).$\footnote{$x$ and $\mu$ here should not to be confused by the spin $x$ axis and the lasers' detunings, respectively. Their meaning should be clear in the context they appear.}

The familiar Jordan-Wigner transformations $\Phi_n = \prod_{l<n}(i\sigma_z^{(l)})\sigma_-^{(n)}$ and $\Phi_n^{\dagger} = \prod_{l<n}(-i\sigma_z^{(l)})\sigma_+^{(n)}$ can be applied to Eq.~(\ref{eq:HSchwingerI}) in order to map the fermionic degrees of freedom to those of a qubit. A unique feature of the lattice Schwinger model with open boundary condition is that the remaining degrees of freedom that are bosonic, namely gauge links and electric field, can be entirely eliminated in favor of new spin-spin interactions. Explicitly, by performing gauge transformations  $\sigma_\pm^{(n)} \to \prod_{l<n} e^{\pm i\theta_l} \sigma_\pm^{(n)}$, and further imposing the Gauss's law $L_n - L_{n-1} = \frac{1}{2}\left[ \sigma_z^{(n)} + (-1)^n \right]$, the Hamiltonian becomes~\cite{Hamer:1997dx, Martinez:2016yna, muschik2017u}
\begin{eqnarray}
&&H = x \sum_{n=1}^{N-1} \left[ \sigma_+^{(n)}\sigma_-^{(n+1)} + \sigma_+^{(n+1)}\sigma_-^{(n)} \right]+
\nonumber\\
&&\hspace{0.1 in} \sum_{n=1}^{N-1} \left[\epsilon_0+\frac{1}{2}\sum_{m=1}^n \left( \sigma_z^{(m)}+(-1)^m\right) \right]^2+\frac{\mu}{2} \sum_{n=1}^N (-1)^n \sigma_Z^{(n)}.
\nonumber\\
\label{eq:HSchwingerII}
\end{eqnarray}
Here, $\epsilon_0$ is the electric field flux into the first lattice site which can be set to zero without loss of generality. To make explicit the mapping of this Hamiltonians to that of the Hamiltonian of the ion-laser system in our proposed scheme, Eq.~(\ref{eq:Heisenberg}), one can note that Eq.~(\ref{eq:HSchwingerII}) can be rewritten as
\begin{eqnarray}
&&H = H^{(xx)}+H^{(yy)}+H^{(zz)}+H^{(z)},
\label{eq:HSchwingerSplit}
\end{eqnarray}
where
\begin{eqnarray}
&&H^{(xx)}=\frac{x}{2} \sum_{n=1}^{N-1} \sigma_x^{(n)}\sigma_x^{(n+1)},
\label{eq:HSchwingerSplitI}
\\
&&H^{(yy)}=\frac{x}{2} \sum_{n=1}^{N-1} \sigma_y^{(n)}\sigma_y^{(n+1)},
\label{eq:HSchwingerSplitII}
\\
&&H^{(zz)}=\frac{1}{2} \sum_{m=1}^{N-2} \sum_{n=m+1}^{N-1} (N-n) \sigma_z^{(m)}\sigma_z^{(n)},
\label{eq:HSchwingerSplitIII}
\\
&&H^{(z)}=\frac{\mu}{2} \sum_{n=1}^N (-1)^n \sigma_z^{(n)}-\frac{1}{2} \sum_{n=1}^{N-1} (n~\text{mod}~2) \sum_{l=1}^n \sigma_z^{(l)}.
\nonumber\\
\label{eq:HSchwingerSplitIV}
\end{eqnarray}
$H^{(xx)}$ and $H^{(yy)}$ represent nearest-neighbor spin-spin interactions and share the same coupling strength. $H^{(zz)}$ is a long-range spin-spin interaction, representing the 1D Coulomb interaction among the charged fermions.
\begin{figure*}[t!]
\includegraphics[scale=0.535]{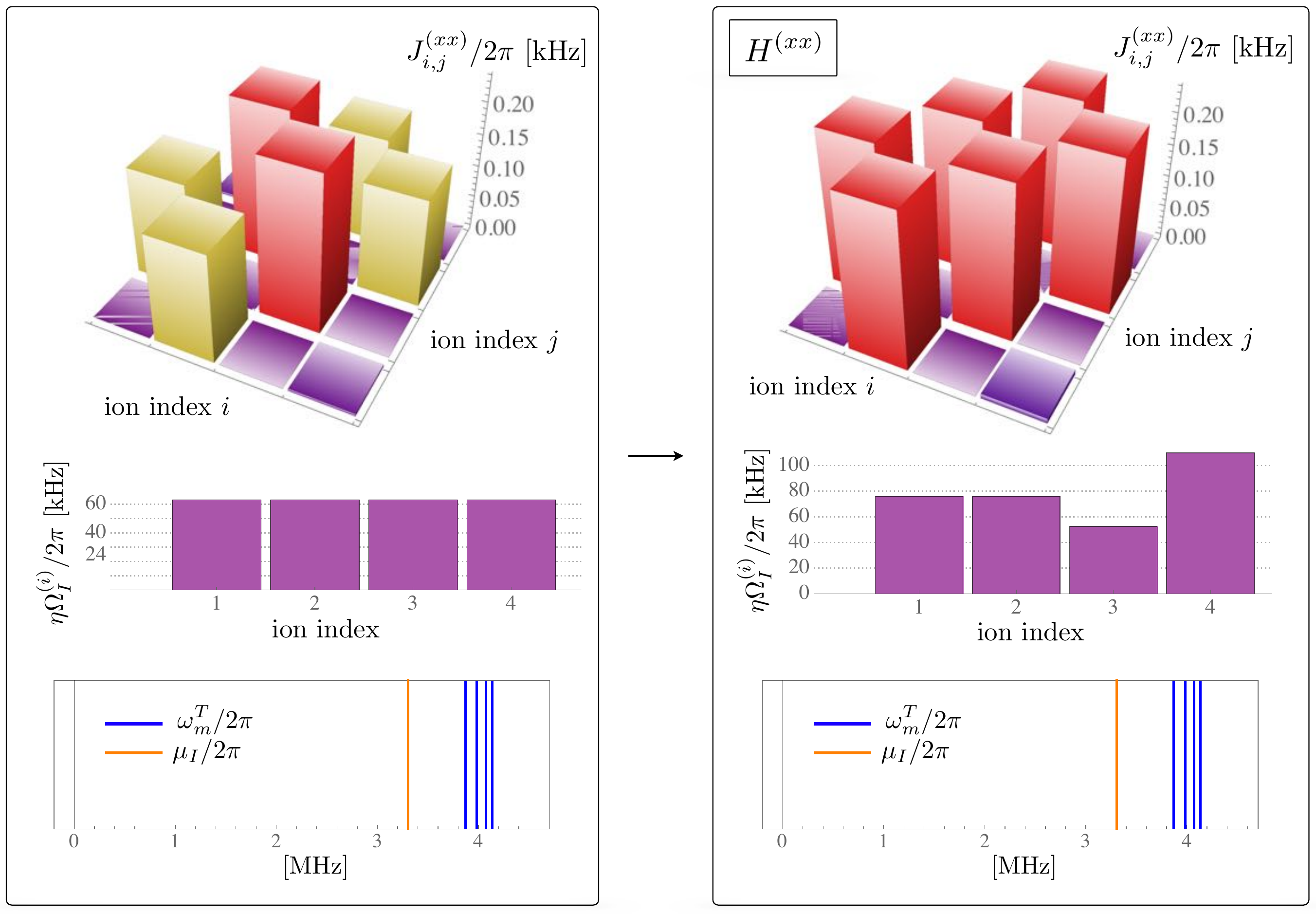}
\caption[.]{
[Left panel] The effective spin-spin coupling matrix $J^{(xx)}$ in Eq.~(\ref{eq:JXXdef}) resulting from pairs of Raman beams addressing 4 individual ions at the Rabi frequency $\Omega^{(i)}$, where $i=1,\cdots,4$. All beams are detuned from the transverse COM mode, $\omega_1^T=2\pi \times 4.135~\text{MHz}$, by the same frequency, $\mu_I - \omega_1^T=-2\pi\times 830$ kHz. The Lamb-Dicke parameter, $\eta$, multiplying the Rabi frequencies in the figure is $\eta=\sqrt{{(\Delta k_I)^2}/{4\pi M\nu^T}} \approx 0.068$. [Right panel] With the same detuning, the Rabi frequencies can be adjusted to match the magnitude of the $J^{(xx)}_{i,j}$ matrix elements for $|j-i|=1$ in the left panel, producing exactly equal magnitude on these elements, in addition to small non-nearest neighbor contributions, as shown in the right panel. Here, the $J^{(xx)}$ matrix is tuned to produce $H^{(xx)}$ of the 4 fermion-site Schwinger model in Eq.~(\ref{eq:HSchwingerSplitI}) with $x=6$. Numerical values associated with this figure are provided in Supplemental Material.
}
\label{fig:xxnnfor4ions}
\end{figure*}

Given the experimental setup presented in the previous section, engineering the Schwinger Hamiltonian for given values of $N$ (which maps directly to the number of ions), $x$ and $\mu$ amounts to finding values of lasers' Rabi frequencies, $\Omega_{I}^{(i)}$, $\Omega_{II}^{(i)}$, and $\Omega_{III}^{(i)}$, and their detunings $\mu_{I}$, $\mu_{II}$, and $\mu_{III}$, as well as $B_z^{(i)}$ values induced by a Stark shift, such that the ion-laser Hamiltonian in Eq.~(\ref{eq:Heisenberg}) is equal to the Schwinger Hamiltonian in Eq.~(\ref{eq:HSchwingerSplit}). This is a well-constrained optimization problem provided that multiple laser frequencies are used with each set of beams each with a corresponding Rabi frequency such that the total number of free parameters, $N\widetilde{n}_{\mu_L}$, is no less than the number of independent nonzero elements in each $J_{i,j}$ coupling matrix, that is $N(N-1)/2$. Here, $\widetilde{n}_{\mu_L}$ is the number of beatnote frequencies on each pair of lasers $L$. Note that this can be achieved with $\widetilde{n}_{\mu_L} \leq N$. It is, however, conceivable that in the first generation of experiments planned, only the amplitude control of Raman beams will be a reality. As a result, we first focus on experimental proposals that do not require a frequency control.
\begin{figure*}[t!]
\includegraphics[scale=0.535]{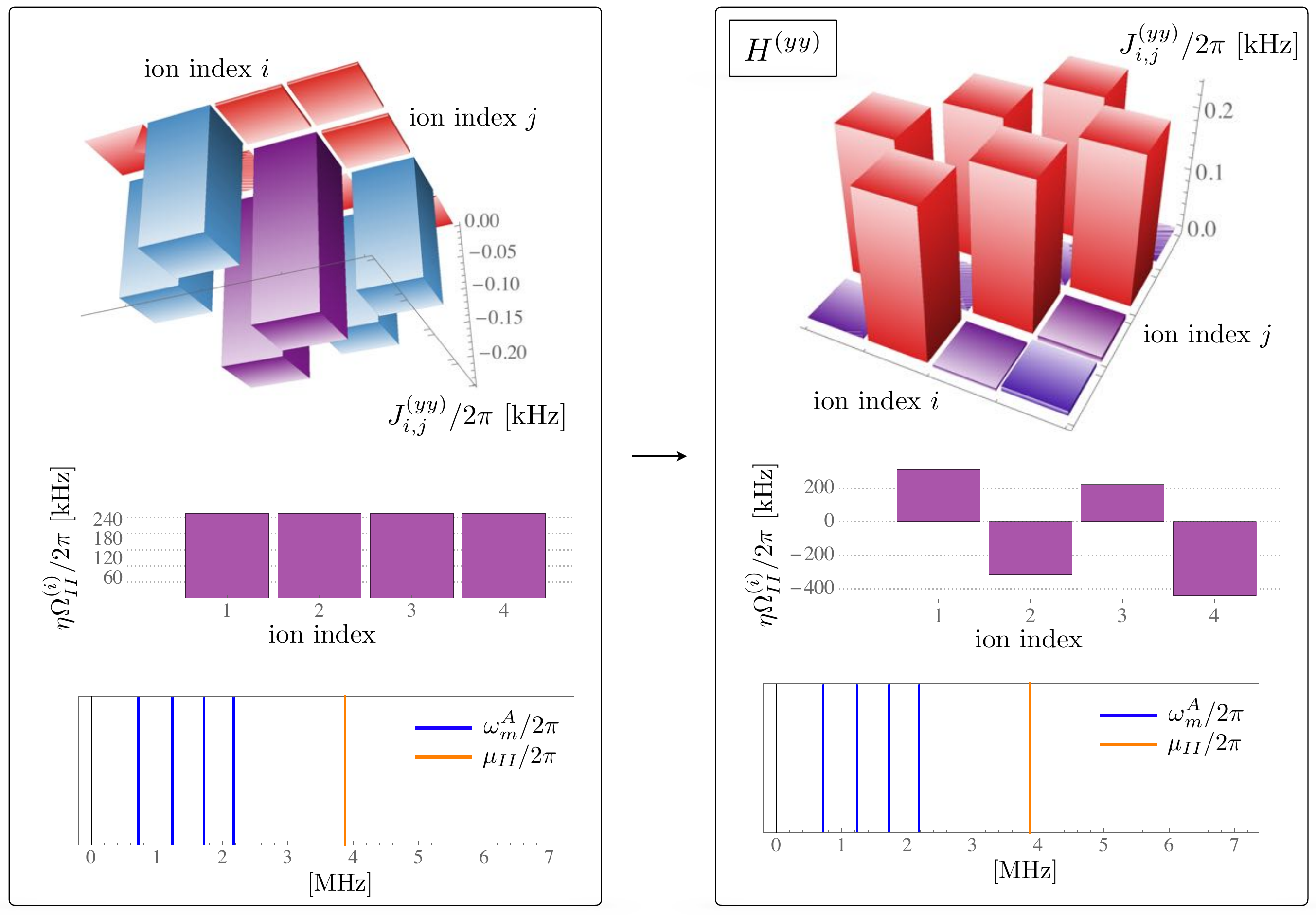}
\caption[.]{
[Left panel] The effective spin-spin coupling matrix $J^{(yy)}$ in Eq.~(\ref{eq:JYYdef}) resulting from pairs of Raman beams addressing 4 individual ions at the Rabi frequency $\Omega^{(i)}$, where $i=1,\cdots,4$. All beams are detuned from the axial COM mode, $\omega_4^A=2\pi\times0.713~\text{MHz}$, by the same frequency, $\mu_{II} - \omega_4^A=2\pi\times 3160$ kHz. The Lamb-Dicke parameter, $\eta$, multiplying the Rabi frequencies in the figure is $\eta=\sqrt{{(\Delta k_{II})^2}/{4\pi M\nu^A}} \approx 0.081$. [Right panel] With the same detuning, the Rabi frequencies can be adjusted to match the magnitude of the $J^{(yy)}_{i,j}$ matrix elements for $|j-i|=1$ in the left panel, producing exactly equal magnitude on these elements, in addition to small $\lesssim3\%$ non-nearest neighbor contributions, as shown in the right panel. Here, the $J^{(yy)}$ matrix is tuned to produce $H^{(yy)}$ of the 4 fermion-site Schwinger model in Eq.~(\ref{eq:HSchwingerSplitII}) with $x=6$. Numerical values associated with this figure are provided in Supplemental Material.}
\label{fig:yynnfor4ions}
\end{figure*}
\begin{figure*}[t!]
\includegraphics[scale=0.535]{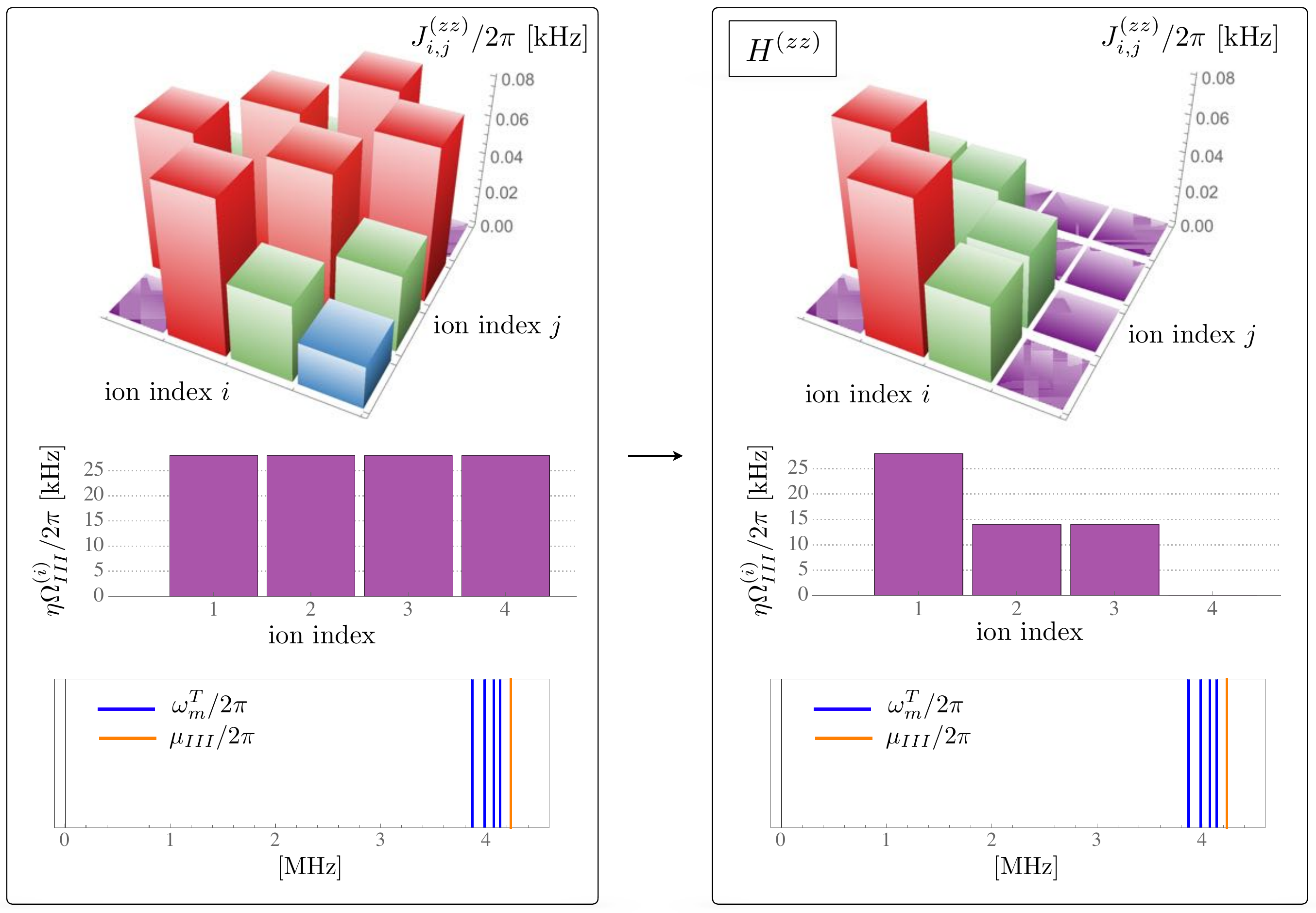}
\caption[.]{
[Left panel] The effective spin-spin coupling matrix $J^{(zz)}$ in Eq.~(\ref{eq:JZZdef}) resulting from pairs of Raman beams addressing 4 individual ions at the Rabi frequency $\Omega^{(i)}$, where $i=1,\cdots,4$. All beams are detuned from the transverse COM mode, $\omega_1^T=2\pi\times4.135~\text{MHz}$, by the same frequency, $\mu_{III} - \omega_1^T=2\pi\times 100$ kHz. The Lamb-Dicke parameter, $\eta$, multiplying the Rabi frequencies in the figure is $\eta=\sqrt{{(\Delta k_{III})^2}/{4\pi M\nu^T}} \approx 0.068$. [Right panel] With the same detuning, the Rabi frequencies can be adjusted so that the $J^{(zz)}$ matrix produces the long-range couplings in $H^{(zz)}$ of the 4 fermion-site Schwinger model in Eq.~(\ref{eq:HSchwingerSplitIII}) with $x=6$. Numerical values associated with this figure are provided in Supplemental Material.}
\label{fig:zzlrfor4ions}
\end{figure*}
\begin{figure*}[t!]
\includegraphics[scale=0.55]{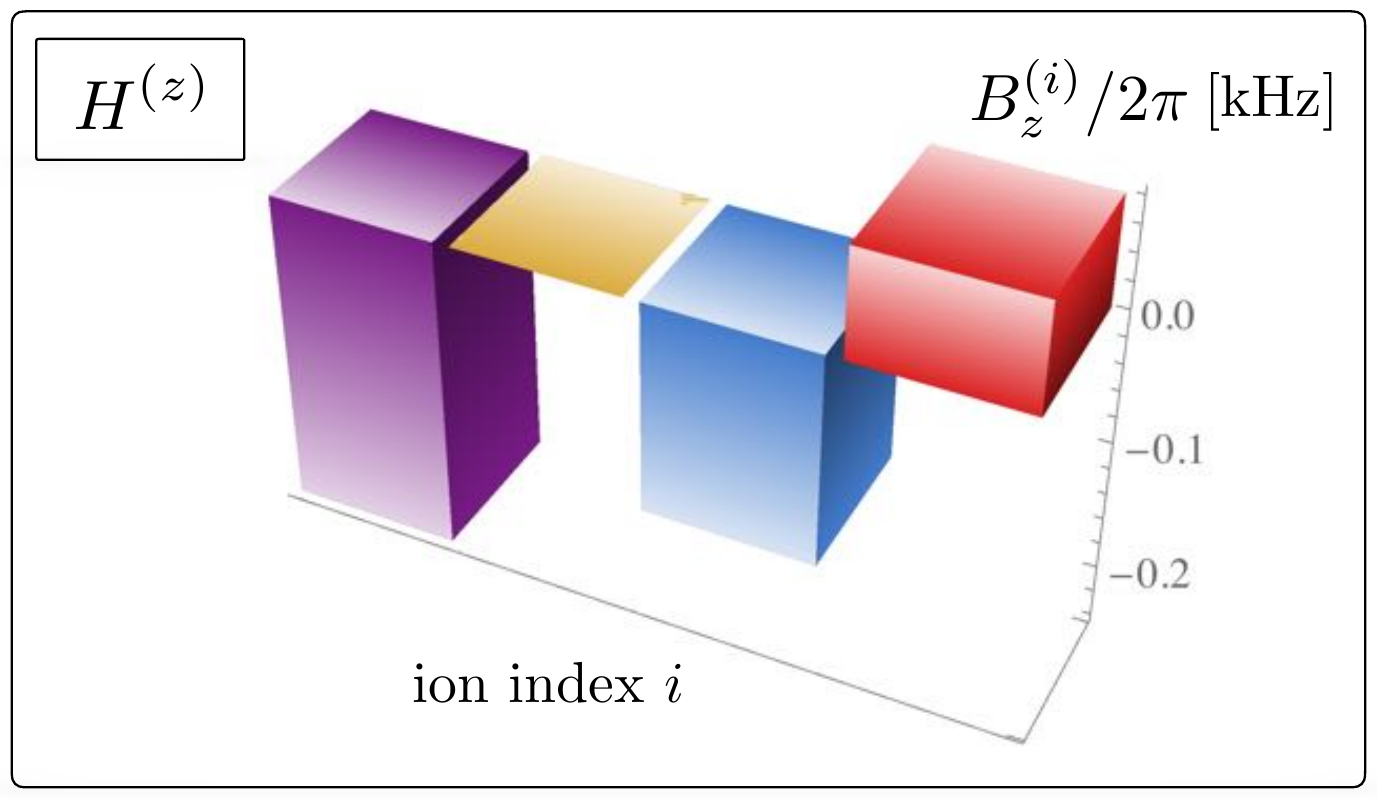}
\caption[.]{The effective magnetic field on each ion, $B_z^{(i)}$, that produces the $H^{(z)}$ Hamiltonian of the Schwinger model, Eq.~(\ref{eq:HSchwingerSplitIV}), for $N=4$ and $\mu=1$. Numerical values associated with this figure are provided in Supplemental Material.}
\label{fig:bzfor4ions}
\end{figure*}
%
\subsection*{A single-detuning and multi-amplitude scheme}
With a single beatnote frequency on each pair of Raman beams, the Schwinger Hamiltonian on small lattices can be realized with good accuracy. For this example, an ion trap consisting of $^{171}\text{Yb}^+$ ions will be considered. The specifications of this system are presented in Appendix~\ref{app:Yb}. Consider the case of $N=4$, and further set the values of the parameters of the Schwinger Hamiltonian to $x=6$ and $\mu=1$. The Hamiltonian $H^{(xx)}$ can be achieved by first noting that a certain detuning from the CM transverse mode with the same amplitude on each ion produces the coupling matrix shown in the left panel of Fig.~\ref{fig:xxnnfor4ions}. This matrix can be systematically turned into a nearest-neighbor form: the slope of the decline in the strength of nearest-neighbor couplings from the center of the chain can be determined, and be used to systematically adjust the Rabi frequencies in such a way that an equal strength is achieved on all $J_{i,j}$ with $|i-j|=1$, as demonstrated in the right panel of Fig.~\ref{fig:xxnnfor4ions}. The most accurate nearest-neighbor Hamiltonian achieved with this procedure presents a $\sim3\%$ contamination on the non-nearest-neighbor elements, and no contamination on the nearest-neighbor elements.

As mentioned in Sec.~\ref{sec:ionphysics}, the $H^{(yy)}$ effective Hamiltonian is chosen to arise from the Raman beams that address the axial modes of the ions. If the transverse modes were to be addressed, the Raman beams would have to be detuned from the modes by the same amount as those for the $H^{(xx)}$ Hamiltonian, as these appear with the same coupling in the Schwinger Hamiltonian. This however would cause the dynamics to deviate from the effective Heisenberg model in Eq.~(\ref{eq:Heisenberg}), given the non-zero commutations between $\widetilde{H}_I$ and $\widetilde{H}_{II}$ in Eqs.~(\ref{eq:HintI}) and~(\ref{eq:HintII}), generating phonon-dependent terms that grow (or decline) linearly with time. Such contaminations are circumvented by producing the $H^{(yy)}$ Hamiltonian with the Raman beams that couple to the axial modes. Note that the axial modes have a very different frequency spectrum compared with the transverse modes. The same procedure as for the $H^{(xx)}$ mapping can be used to find the values of the laser beatnote and Rabi frequencies that generate a nearest-neighbor interaction with these modes, see Fig.~\ref{fig:yynnfor4ions}. As discussed at the end of Sec.~\ref{sec:ionphysics}, a critical check is to ensure the higher-sideband contributions to the applied Molmer-Sorenson scheme are not significant given the low normal-mode frequencies in the axial direction, and given the laser frequencies applied. It can be shown that the largest contribution from these higher-order sidebands is only a few percent of the contribution from the first sideband, and will be ignored in the current proposal.

An effective $H^{(zz)}$ Hamiltonian that matches that of the Schwinger model can be achieved with a single beatnote frequency, and by addressing the other set of transverse normal modes of the ions. Here, the values shown in Fig.~\ref{fig:zzlrfor4ions} allow the $J_{i,j}$ coupling to be tuned to the desired values with below-percent accuracy. However, in contrast with the case of nearest-neighbor Hamiltonians, the procedure that finds the adjusted Rabi frequencies for $H^{(zz)}$ is not systematic, making it challenging to generalize such an ad hoc tuning procedure to a higher number of ions. Finally, an effective $H^{(z)}$ Hamiltonian can be induced using $N$ sets of Raman beams with their Stark shift tuned to reproduce $H^{(z)}$ of the Schwinger Hamiltonian in Eq.~(\ref{eq:HSchwingerSplitIV}). The values of the effective magnetic fields that are required given the chosen parameters of the model are depicted in Fig.~\ref{fig:bzfor4ions}.

It is crucial to verify that the laser parameters found in such a mapping do not violate the conditions enumerated in the previous section, and the true dynamics is that dictated by the effective Heisenberg Hamiltonian in Eq.~(\ref{eq:Heisenberg}). This check can be done by a numerical evaluation of all contributions to the exponent of the evolution operator in Eq.~(\ref{eq:UAll}), up to and including $\mathcal{O}(\eta^2,\eta B)$. Here, we assume that the experiment can be initiated in a state with zero phonon occupation in all modes. The results of this investigation are shown in Fig.~\ref{fig:allcontbion1for4ions} of Appendix~\ref{app:numerics} for the first ion, and in Supplemental Material for the rest of the ions. As shown, the dominant source of error is related to the nonzero commutations of $H_B$ in Eq.~(\ref{eq:HB}) with $\widetilde{H}_I$ and $\widetilde{H}_{II}$ in Eqs.~(\ref{eq:HintI}) and~(\ref{eq:HintII}), introducing effective magnetic fields along the $x$ and $y$ spin axes. These are a small fraction of the desired field along the $z$ direction, but are however dependent upon the phonon occupation in the system.

\begin{figure*}[t!]
\includegraphics[scale=0.652]{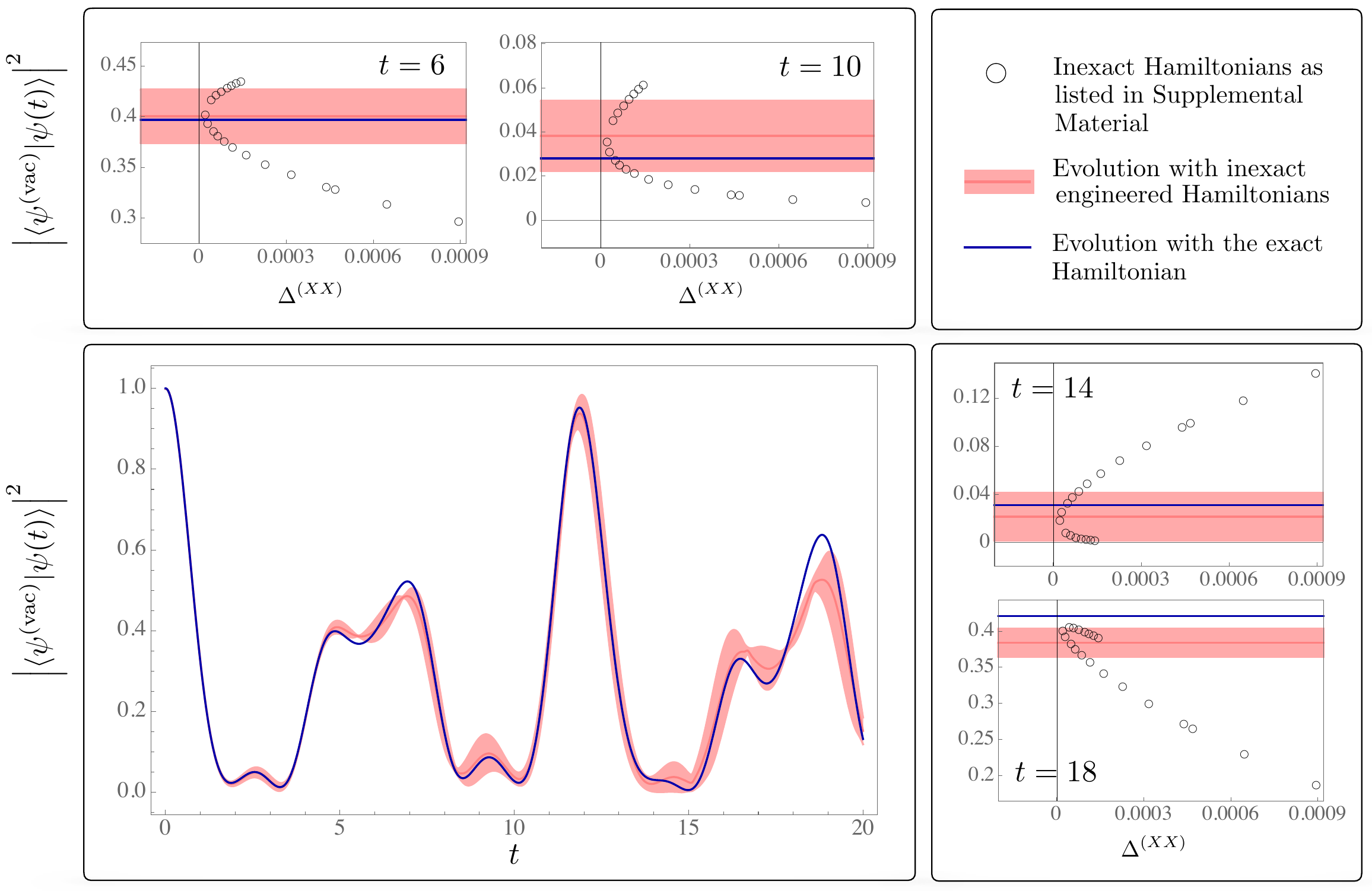}
\caption[.]{Time evolution of the state $|\psi^{(\text{vac})} \rangle = |\downarrow\uparrow\downarrow\uparrow\rangle$ corresponding to the strongly-interacting vacuum of the 4 fermion-site lattice Schwinger model with $x=0.6$ and $\mu=0.1$. Open circles in the upper and left plots are the values of VPA at select times with 20 inexact $H^{(xx)}(=H^{(yy)})$ Hamiltonians, as listed in Supplemental Material. The 9 data points that satisfy $\Delta^{(xx)} \equiv {J_{1,3}^{(xx)}}^2+{J_{1,4}^{(xx)}}^2+{J_{2,4}^{(xx)}}^2 \leq 10^{-4}$ are chosen to define central values (dark-pink lines) and uncertainties (pink bands) on the VPA, and are compared with the exact expectations (blue lines). The plot in the lower-left panel represents the exact time evolution of vacuum (blue curve) compared with the central value (dark-pink curve) and uncertainty (pink band) on the VPA obtained from 9 Hamiltonians that give rise to $\Delta^{(xx)} \leq 10^{-4}$. Numerical values associated with these plots are provided in Supplemental Material.}
\label{fig:schwinger}
\end{figure*}

Hamiltonians of the lattice Schwinger model for a larger number of fermion sites can be shown to be accessible through the single-frequency and multi-amplitude scheme described, but deviations from the exact Hamiltonian can be significant. For $N=10$ and the nearest-neighbor Hamiltonian with transverse modes, the best parameters found give rise to errors as high as $\sim20\%$ in the non-nearest-neighbor elements. To investigate the effect of inexact Hamiltonians on the dynamics of the Schwinger model, we have studied a lattice Schwinger model with $N=4$, $x=0.6$ and $\mu=0.1$ with exact engineered Hamiltonians $H^{(zz)}$ and $H^{(z)}$ but with a nearest-neighbor Hamiltonian $H^{(xx)}(=H^{(yy)})$ that differs from the exact form via nonzero non-nearest-neighbor elements. Twenty such Hamiltonians are considered, as listed in Supplemental Material, with errors on the non-nearest-neighbor elements in the range $\sim 3\%-18\%$. The evolution of the 4 fermion-site Schwinger model is then considered. The quantity of interest here is the vacuum persistence amplitude (VPA), defined as the (square) of the overlap of the state of the system at time $t$, $|\psi(t)\rangle$ with the ``vacuum'' (a state in the physical sector of the theory with no net electron-positron pair), $|\psi^{(\text{vac)}}\rangle$. This quantity is plotted for select times in the smaller panels of Fig.~\ref{fig:schwinger} for all the twenty inexact Hamiltonians used in the evolution. A procedure is described to estimate a mean and uncertainty band from the most accurate Hamiltonians employed. Nonetheless, as is seen in the larger panel of the figure, during certain times, the estimate of VPA deviates significantly from the expected result, and this feature is amplified at longer times.

This observation promotes adopting a multi-frequency and multi-amplitude scheme,\footnote{We use the term frequency for the beatnote frequency of the Raman beams unless it is identified as otherwise. A multi-frequency scheme, therefore, refers to when multiple beatnote frequencies are used, while a multi-amplitude scheme refers to when multiple Rabi frequencies are applied.} as proposed previously in Ref.~\cite{korenblit2012quantum} in the context of quantum simulation of the Ising model on two-dimensional lattices. With this scheme, mapping of the effective Hamiltonian of the ion-laser system to that of the Schwinger model can be achieved with unprecedented accuracy, as is shown in the following.
\begin{figure*}[t!]
\includegraphics[scale=0.65]{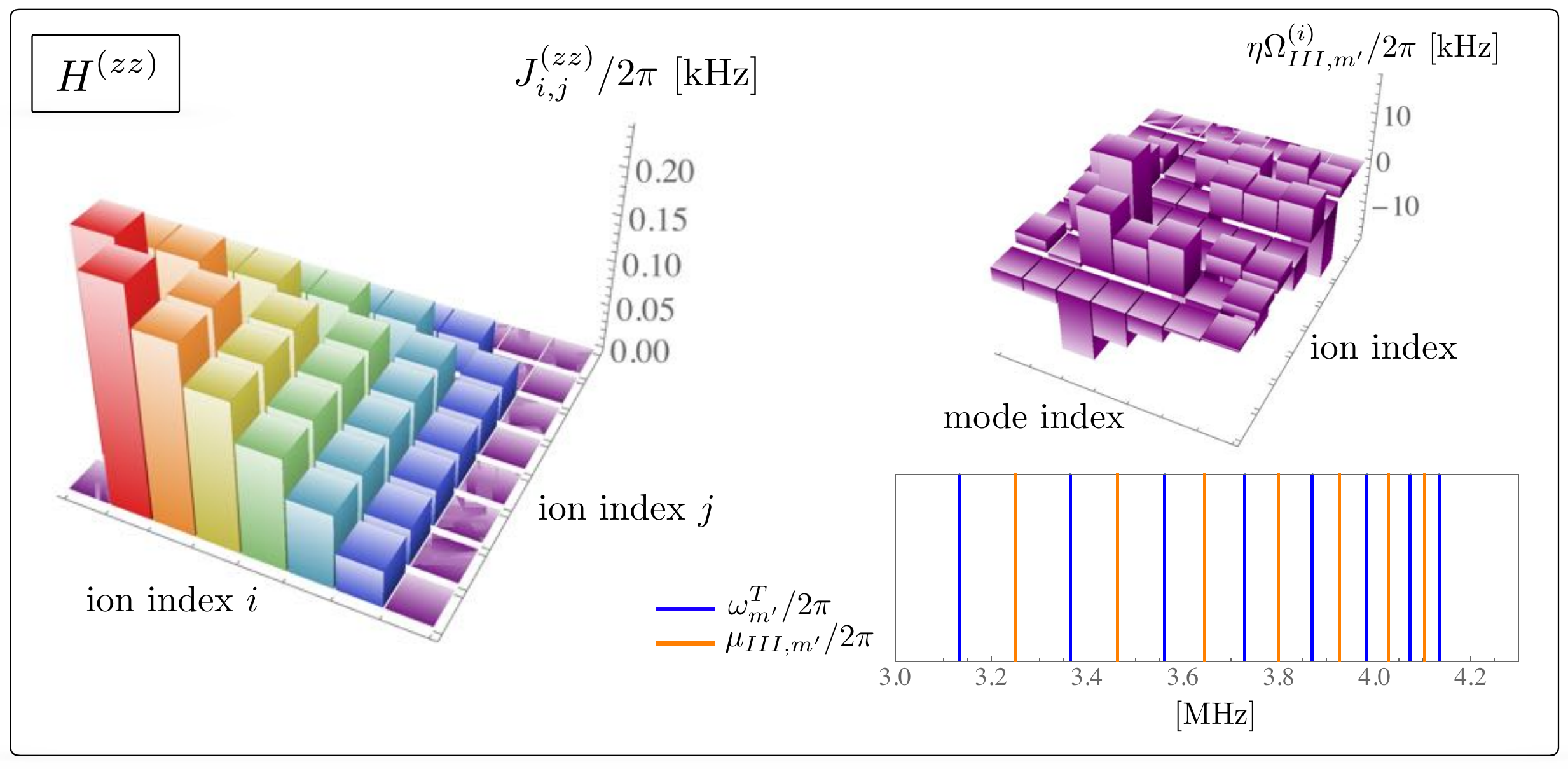}
\caption[.]{
The effective spin-spin coupling matrix $J^{(zz)}$ in Eq.~(\ref{eq:JZZdef}) resulting from multiple pairs of Raman beams addressing $N=8$ individual ions at the Rabi frequency $\Omega_{III,m^\prime}^{(i)}$, where $i=1,\cdots,8$ and $m^\prime=1,\cdots,7$. The pairs of beams addressed at ion $i$ are detuned from the transverse COM mode by 7 different frequencies, $\mu_{III,m^\prime}=\omega_{m^\prime}^T+f_s(\omega_{m^\prime}^T-\omega_{m^\prime+1}^T)$ with $f_s=-0.5$, as denoted in the lower-right of the panel. The Lamb-Dicke parameter, $\eta$, multiplying the Rabi frequencies in the figure is $\eta=\sqrt{{(\Delta k)^2}/{4\pi M\nu^T}} \approx 0.068$. Here, the $J^{(zz)}$ matrix is tuned to produce $H^{(zz)}$ of the 8 fermion-sites Schwinger model in Eq.~(\ref{eq:HSchwingerSplitIII}). Numerical values associated with this figure are provided in Supplemental Material.
}
\label{fig:zzlrfor8ions}
\end{figure*}
%
\subsection*{A multi-frequency and multi-amplitude scheme}
The extension of the formalism presented in Sec.~\ref{sec:ionphysics} to a multi-frequency scheme is straightforward. For example, the effective spin-spin coupling engineered by Raman pairs $I$ generalizes to
\begin{eqnarray}
J_{i,j}^{(xx)}=\sum_{m^\prime=1}^{\widetilde{n}_{\mu_I}}\Omega_{I,m^\prime}^{(i)}\Omega_{I,m^\prime}^{(j)} R_I \sum_{m=1}^N \frac{b_{m}^{(i)}b_{m}^{(j)}}{{\mu_{I,m^\prime}}^2-{\omega_m^T}^2},
\label{eq:JijGeneral}
\end{eqnarray}
where $\widetilde{n}_{\mu_I}$ is the number of beatnote frequencies, and where each detuning $\mu_{I,m^\prime}$ is associated with the Rabi frequency $\Omega_{I,m^\prime}^{(i)}$.\footnote{We remind that the effective spin-spin Hamiltonian arises from a bichromatic pair of Raman beams, one detuned by $-\mu_{I,m^\prime}$ (red detuned) and one by $\mu_{I,m^\prime}$ (blue detuned) from the carrier transition, see discussions after Eq.~(\ref{eq:HintIII}).} Similarly, the $J_{i,j}^{(yy)}$ and $J_{i,j}^{(zz)}$ coupling matrices can be obtained by replacements $\mu_{II} \to \mu_{II,m^\prime}$, $\Omega_{II}^{(i)} \to \Omega_{II,m^\prime}^{(i)}$, $\mu_{III} \to \mu_{III,m^\prime}$, and $\Omega_{III}^{(i)} \to \Omega_{III,m^\prime}^{(i)}$, where a summation over $m^\prime$ is assumed. For $J_{i,j}^{(yy)}$, one must replace $\omega_m^T$ with $\omega_m^A$. More generally, the full time evolution operator in Eq.~(\ref{eq:UAll}) can be constructed by performing the changes described in the ion-laser Hamiltonians in Eqs.~(\ref{eq:HintI}-\ref{eq:HintIII}). This introduces additional off-resonant terms that would scale as the number of beatnote frequencies introduced. One therefore needs to ensure that the cumulative effect of such terms remain negligible compared with the desired effective Heisenberg Hamiltonian.

Fig.~\ref{fig:zzlrfor8ions} demonstrates the success of this scheme in an accurate generation of the long-range part of the Schwinger Hamiltonian, $H^{(zz)}$, for the case of $N=8$ ions. Here, the corresponding optimization problem is solved, and the desired effective spin-spin Hamiltonian is achieved with errors that are comparable with the machine precision. The laser frequencies are fixed such that $\mu_{I,m^\prime}=\omega_{m^\prime}^T+f_s(\omega_{m^\prime}^T-\omega_{m^\prime+1}^T)$, with $f_s=-0.5$, and where $m^\prime$ runs from 1 to $\widetilde{n}_{III}=7$, see the lower-right plot of Fig.~\ref{fig:zzlrfor8ions}.\footnote{In the convention of this work, the normal mode frequencies are ordered in a set from the highest value to the lowest value. Therefore for the axial mode, $\omega_N^A$ denotes the CM mode, while for the transverse mode,  the CM mode is $\omega_1^T$. Because of this convention, the normal-mode eigenvectors $b_m^{(i)}$ must be ordered accordingly for the transverse and axial modes.} The corresponding Rabi frequencies at the location of each ion are plotted in the upper-right plot of Fig.~\ref{fig:zzlrfor8ions}. As is seen, a perfect agreement between $J_{i,j}^{(zz)}$ and that in the Schwinger model with $x=6$ and $\mu=1$ is achieved. The reason for choosing a large value of the coupling $x$ in the original theory is to minimize the error to the effective Heisenberg Hamiltonian due to the unbounded contributions arising from the commutations of the $B_z$ Hamiltonian in Eq.~(\ref{eq:HB}) and $\widetilde{H}_I$ and $\widetilde{H}_{II}$ in Eqs.~(\ref{eq:HintI}) and~(\ref{eq:HintIII}). Note that the desired effective $B_z$ field in the Schwinger Hamiltonian grows with $N$ even in the limit $\mu=0$. Hence, in order to keep the undesired contribution small compared with the effective Hamiltonian, the strength of the nearest-neighbor terms is taken to be stronger by setting $x=6$. As is shown in Appendix~\ref{app:numerics} for the first ion, and in Supplemental Material for the rest of the ions, all the contributions to the exponent in the full time-evolution operator (up to the order considered) are small (and mostly bounded) compared with those that constitute the Hamiltonian of the Schwinger model. The laser parameters for a nearly exact engineering of $H^{(xx)}$, $H^{(yy)}$, and $H^{(z)}$ are shown in Figs.~\ref{fig:xxnnfor8ions}-\ref{fig:bzfor8ions} of Appendix~\ref{app:N8plots}. It must be noted that the optimization problem in all cases is solved under two constraints: i) the sum of Rabi frequencies at the location of each ion is less than or equal to 2~MHz, ii) the contribution to the full evolution from the first-order terms, those proportional to coefficients $\alpha_{i,m}^{(x)}$, $\alpha_{i,m}^{(y)}$, and $\alpha_{i,m}^{(z)}$ in Eqs.~(\ref{eq:phiX}-\ref{eq:phiZ}), remains below $0.5$ at several random times up to 1~ms.

To summarize, we have provided detailed experimental protocols for a fully analog simulation of the Schwinger model for given parameters with i) a scheme that requires only individual amplitude and phase control of the laser beams and engineers an approximate Schwinger Hamiltonian, and ii) a scheme that takes advantage of individual amplitude, phase and frequency control and engineers the desired Hamiltonian with great accuracy (up to errors associated with the difference between the full ion-laser evolution and the effective Heisenberg model, which are nonetheless assured to remain negligible in the schemes proposed). It is clear that the second scheme can be easily applied to any number of ions at the cost of introducing a multitude of laser frequencies, the number of which grows with the number of ions. This can be already achieved with current technologies for up to $\sim30$ ions, and most importantly is scalable, as it involves a linear growth in the complexity of the classical control hardware of the experiment.

In the following, other examples of LGTs whose dynamics can be mapped onto a spin-$\frac{1}{2}$ system will be discussed. The goal is to only point out the potential of an ion-trap quantum simulator in addressing more complex spin systems by providing examples of relevant gauge theories. Explicit scenarios for given ion-trap architectures are straightforward to obtain, following optimization strategies presented for the case of the Schwinger model.

\section{Analog simulations of systems in higher dimensions with a 1D chain of ions
\label{sec:higherd}}
\noindent
With a generic Heisenberg model and an effective magnetic field engineered in Sec.~\ref{sec:ionphysics}, it is clear that a wide range of couplings among spins can be tailored, as was demonstrated for the case of the Schwinger model. In particular, as seen in Sec.~\ref{sec:schwinger}, the $H^{(\alpha\alpha)}$ with $\alpha=x,y,z$ does not have to be necessarily nearest neighbor or of any particular form, as the multi-frequency, multi-amplitude scheme of this work allows an arbitrary $J_{i,j}$ to be produced. This observation implies that spin systems in higher spatial dimensions can be engineered as well, as was also noted in Ref.~\cite{korenblit2012quantum}. One only needs to map the points on a 2D or 3D lattice to a linear chain of ions along with their corresponding couplings. Of course, with a fixed number of ions in a given experiment, this means that the finite-size effects in the dynamics of the system under study will be larger, as e.g., in the case of square and cubic lattices the spatial extent of the system will be $N^{1/2}$ and $N^{1/3}$, respectively. Nevertheless, this possibility implies that a linear quantum system can be used as a platform for analog simulations of theories in any dimension, bringing the versatility of such an analog platform closer to its digital counterpart.

\begin{figure*}[t!]
\includegraphics[scale=0.55]{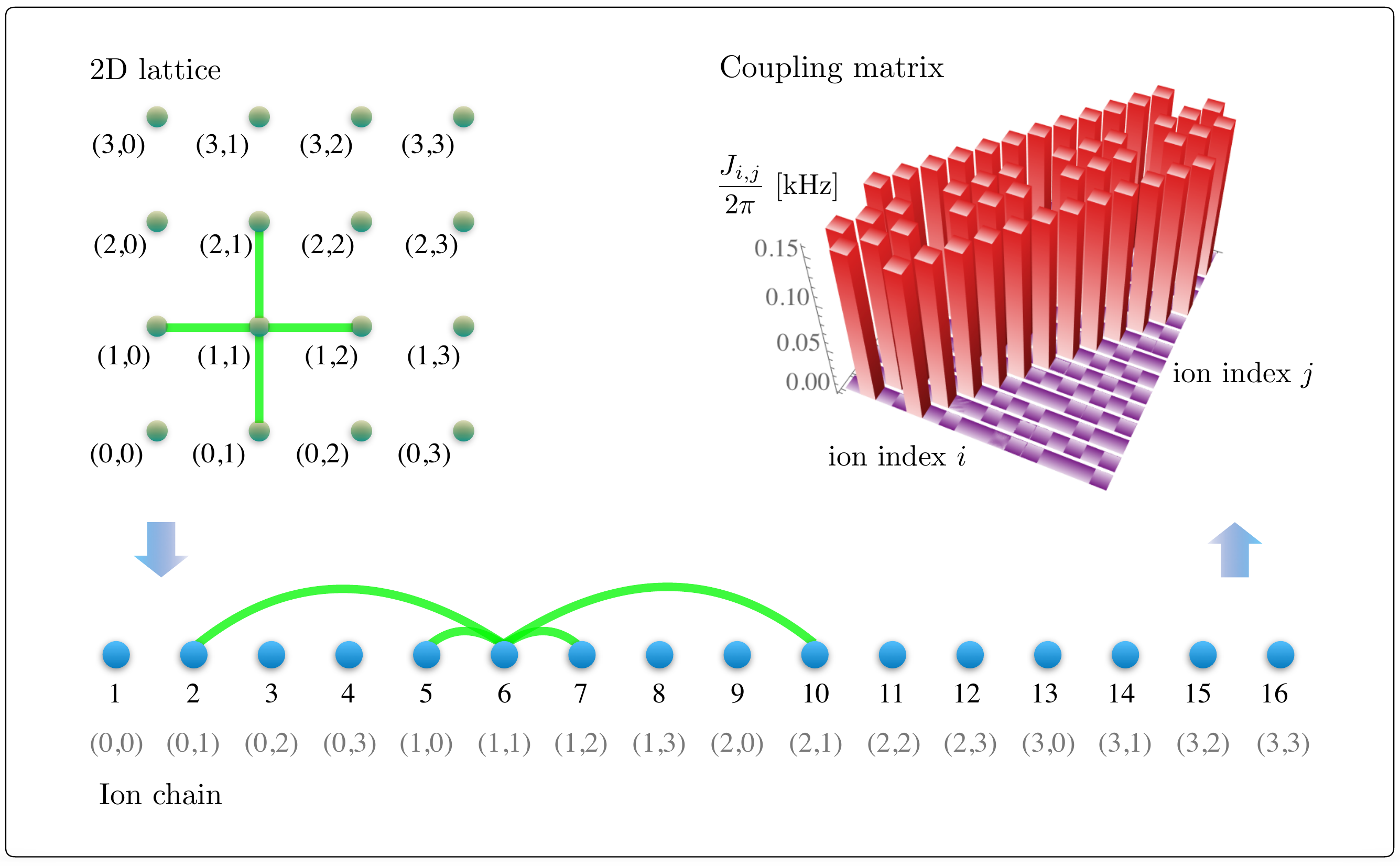}
\caption[.]{The upper left panel shows a $4 \times 4$ lattice of spins ($s=\frac{1}{2}$) with nearest-neighbor interactions, corresponding to the $\sigma_x\otimes\sigma_x$ (or equivalently $\sigma_y\otimes\sigma_y$) interactions in the Hamiltonian in Eq.~(\ref{eq:H-CS-transformed}) with $\bm{n}=(n_x,n_y)$, where $n_x$ ($n_y$) runs from $0$ to $3$, and where an open boundary condition is adopted. The nearest-neighbor interactions of a select site are depicted in green links. This 2D configuration can be mapped to a 1D chain of ions, as shown in the lower panel, along with the couplings of the select site in the new configuration. The obtained 1D coupling matrix $J_{i,j}$ is shown in the upper right panel.
}
\label{fig:nn2dlattice}
\end{figure*}
\subsection{2+1D Abelian Chern-Simons theory coupled to fermions}
As an example of an interesting field theory in $2+1$D, consider the Chern-Simons theory coupled to fermions. This theory is of broad impact on a range of problems in theoretical physics, from  the theory of the integer and fractional quantum Hall effects to knot theory and parity anomalies in quantum field theory, see Ref.~\cite{fradkin_2013} for a review. Since the theory is topological in the continuum, the construction of a discretized counterpart of the theory turned out to be non-trivial as a lattice has explicit reference to a given coordinate system and metric. However, it has been shown~\cite{fradkin1989jordan,sun2015discretized} that one can still formulate a U(1) LGT that retains gauge invariance on arbitrary 2D planar graphs, has no local excitations (hence is topological) and in the long-wavelength limit approaches the Chern-Simons theory in the continuum. As is discussed in Ref.~\cite{sun2015discretized}, a lattice formulation of the Chern-Simons theory is invaluable in investigations of fractional Chern insulators that occur in given lattice geometries. As a result, it is interesting to ask if a quantum-simulation protocol for this theory can be devised on the simulating platform of this work.

A known result~\cite{fradkin1989jordan} in the context of the generalized Jordan-Wigner transformation in higher dimensions is Fradkin's proof of equivalence between the spin-$\frac{1}{2}$ XY model on a 2D Bravais lattice and a Chern-Simons theory in 2+1D coupled to fermions, provided that the strength of the Chern-Simons' term in the Lagrangian density:
\begin{eqnarray}
\mathcal{L}_{\text{CS}}&=&a^\dagger(x)iD_0a(x)-\sum_{j=1,2}\left[a^\dagger(x) e^{iA_j(x)}a(x+\hat{\bm{n}}_j)+\right.
\nonumber\\
&&\hspace{2.25 cm} \left. \text{h.c.}\right]-\frac{\theta}{4}\epsilon^{\mu\nu\lambda}A_\mu(x) F_{\nu\lambda}(x)
\label{eq:L-CS}
\end{eqnarray}
is $\theta=\frac{1}{2\pi}$~\cite{fradkin1989jordan}. Here, time is assumed to be continuous while spatial coordinates are defined on a square lattice, i.e., $x=(t,\bm{n})$ where $\bm{n}$ is a vector whose components are integer multiples of the lattice spacing.\footnote{For a general formulation on  2D planar lattices, see Refs.~\cite{sun2015discretized, fu2018properties}.} $\mu,\nu=0,1,2$ with the zeroth direction being the time direction, $a$ is a complex spinless fermion field, $A_{\mu}$ is the gauge field, $D_\mu=\partial_\mu-iA_\mu$ is the covariant derivative, $F_{\mu\nu}$ is the field-strength tensor: $F_{\mu\nu}=\partial_\mu A_\nu-\partial_\nu A_\mu$, and $\epsilon^{\mu\nu\lambda}$ is the Levi-Civita symbol. Note that the $A_0$ field does not have any dynamics and can be set equal to zero with the choice of a temporal gauge. The physical sector of the theory, i.e., states that satisfy the Gauss's law, can be identified from the condition $\frac{\delta S}{\delta A_0}=0$, where $S$ is the action. These states then correspond to those for which $a^\dagger(x) a(x) - \theta \epsilon_{ij} \left[A_j(x+\hat{\bm{n}}_i)-A_j(x)\right]=0$. It is also clear that the Hamiltonian of the theory vanishes in the absence of matter fields, which is a desired feature of the topological theory. In the presence of matter fields, i.e., the Hamiltonian corresponding to Eq.~(\ref{eq:L-CS}) is 
\begin{eqnarray}
H_{\text{CS}}=\sum_{\bm{n}}\sum_{j=1,2}\left[a^\dagger(\bm{n}) e^{iA_j(\bm{n})}a(\bm{n}+\hat{\bm{\bm{n}}}_j)+\text{h.c.}\right].
\label{eq:H-CS}
\end{eqnarray}
Note that the time dependence of the fields is now implicit considering the Hamiltonian equations of motion. As is shown in Ref.~\cite{fradkin1989jordan}, the gauge links can be eliminated from the Hamiltonian with the use of Gauss's law, at the cost of changing the equal-time commutation relation of fermions. This is in fact a great advantage since when $\theta=\frac{1}{2\pi}$ (or in general when $\frac{1}{2\theta}$ is an odd multiple of $\pi$), the new commutation relations are those of hardcore bosons, i.e., the spin-$\frac{1}{2}$ matrices. As a result, this procedure can be realized as a 2D generalization of the familiar Jordan-Wigner transformation. Explicitly, by performing the transformations $a \to e^{i\mathbb{\phi}} a \equiv \tilde{a}$ and $a^\dagger \to a^\dagger e^{-i\mathbb{\phi}} \equiv \tilde{a}^\dagger$, where $A_j(\bm{n}) \equiv \phi_j(\bm{n}+\hat{\bm{n}}_i)-\phi_j(\bm{n})$, one arrives at
\begin{eqnarray}
H_{\text{CS}}=\sum_{\bm{n}}\sum_{j=1,2}\left[\sigma_+^{(\bm{n})} \sigma_-^{(\bm{n}+\hat{\bm{\bm{n}}}_j)}+\text{h.c.}\right],
\label{eq:H-CS-transformed}
\end{eqnarray}
where the following identifications are assumed: $\sigma_+^{(\bm{n})} = \tilde{a}^\dagger(\bm{n})$, $\sigma_-^{(\bm{n})} = \tilde{a}(\bm{n})$, and $\sigma_z^{(\bm{n})}=1-2a^\dagger(\bm{n})a(\bm{n})$. Eq.~(\ref{eq:H-CS-transformed}) clearly corresponds to an XY spin model. Note that a parameter $h$ could be introduced to control the magnitude of the hopping term in the Hamiltonian.

To perform an analog simulation of such a 2D XY model within the scheme presented in Sec.~\ref{sec:ionphysics} requires optimizing a $(J_{i,j} \equiv)~J^{(XX)}_{i,j}=J^{(YY)}_{i,j}$ matrix by performing a multi-frequency, multi-amplitude Molmer-Sorenson scheme using the transverse and axial normal modes of motion. For a $4 \times 4$ lattice in the target theory, a system of $N=16$ ions can be used as is shown in Fig.~\ref{fig:nn2dlattice}, along with the required $J_{i,j}$ matrix. Obtaining the laser frequencies and amplitudes is a straightforward optimization process, as detailed in the previous section, and in fact machine precision accuracy can be achieved, as demonstrated in Ref.~\cite{korenblit2012quantum} for similar geometry and coupling profiles. Finally, we should remark that the full Hamiltonian in such a 2+1D Abelian LGT must include the energy stored in electric and magnetic fields, giving rise to the Maxwell-Chern-Simons theory~\cite{jackiw1990self, diamantini1993topological}.\footnote{See also Ref.~\cite{caspar2016doubled} for discussions regarding a non-Abelian case, the Yang-Mills-Chern-Simons theory.} Aside from the question of what is the proper formulation of a discretized Maxwell-Chern-Simons theory, one needs to account for the full dynamics of the gauge fields by mapping them to those in an ion-trap quantum-simulation platform, which is beyond the scope of the present work.

\subsection{2+1D pure $Z_2$ lattice gauge theory}
\begin{figure*}[t!]
\includegraphics[scale=0.55]{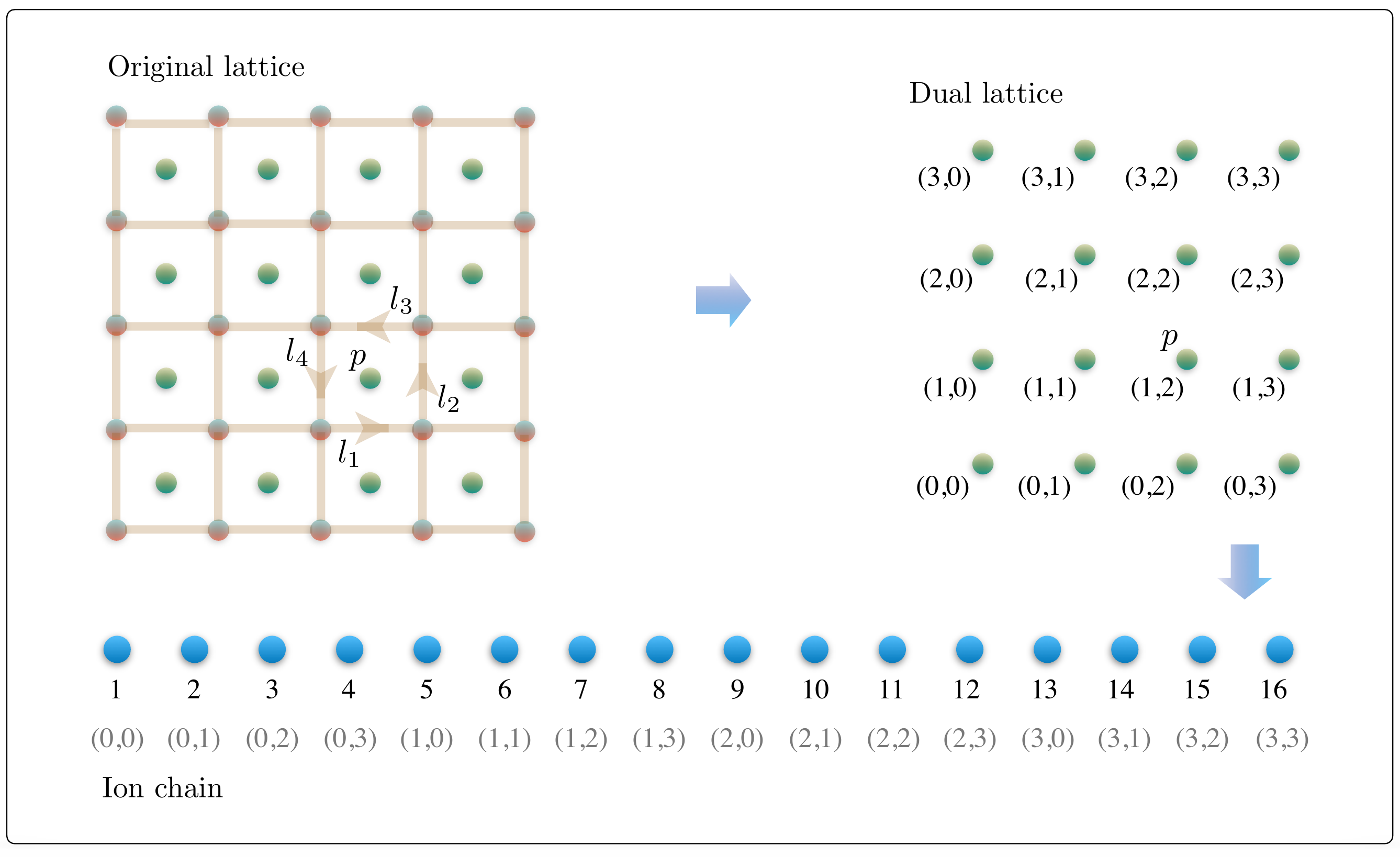}
\caption[.]{The upper-left panel shows a $5 \times 5$ spatial lattice corresponding to the $Z_2$ Hamiltonian in Eq.~(\ref{eq:HZ2}). An open boundary condition is adopted, and a select plaquette term in the Hamiltonian is shown. The center of the plaquettes defines the sites of a dual lattice, as depicted by the green points, and are separately shown in the right panel. Such a 2D configuration corresponds to the Ising Hamiltonian in Eq.~(\ref{eq:H2DIsing}), which can now be mapped to a 1D chain of ions, as shown in the lower panel of the figure.
}
\label{fig:z2ising}
\end{figure*}
$Z_N$ gauge theories are discrete Abelian gauge theories that given their simple underlying symmetry have long served as a testbed for gaining deeper perspectives on gauge theories. Despite their simple structure, they can have non-trivial phase diagrams exhibiting e.g., a confining phase. In fact, since $Z_3$ is the center of the SU(3) group, the confinement in the Yang-Mills theory is attributed to the $Z_3$ symmetry. These gauge theories have been the focus of numerous theoretical and experimental proposals for quantum simulation of gauge theories, in particular using neutral atoms in optical lattices~\cite{Wiese:2013uua,Barbiero:2018wui,schweizer2019floquet}. An interesting feature of $Z_N$ is its duality with spin models. This connection has been developed over decades~\cite{kogut1979introduction}, starting from Wegner's demonstration of such a duality for the case of a $Z_2$ LGT~\cite{Wegner:1984qt}, and has inspired similar duality constructions for non-Abelian gauge theories such as SU(N)~\cite{mathur2016lattice}. Further, recent work has suggested that the 4D $Z_2$ LGT provides a complete model for all classical spin models and all Abelian discrete LGTs~\cite{de2009unifying,de2010mapping}.

The example that will be presented here is a 2+1D $Z_2$ LGT that is dual to a 2D Ising model, and is therefore amenable to the quantum simulation protocol of this work. The Hamiltonian of the 2+1D $Z_2$ LGT can be expressed with a pair of conjugate spin operators $\{\sigma_x(l),\sigma_z(l)\}$, where $\sigma_x(l)=e^{i \pi E(l)}$ and $\sigma_z(l)=e^{iA(l)}$. Here, $l$ denotes a link on the 2D spatial lattice, $A(l)$ is the gauge field evaluated on link $l$ with $A(l)=\{0,\pi\}$. $E(l)$ is the corresponding ``electric field'' with $E(l)=\{0,1\}$. Note that in order to keep the presentation simple, we have not used bold-faced quantities for the two-dimensional vectors $A(l)$ and $E(l)$, as their directionality on the 2D plane is implicit from the directionality of the link arguments. The lattice Hamiltonian of such a pure gauge theory consists of ``electric'' and ``magnetic'' terms:
\begin{eqnarray}
H_{\text{2+1D }Z_2}&=&- \sum_l \sigma_x(l) - \lambda \sum_p \sigma_z(l_1)\sigma_z(l_2)\sigma_z(l_3)\sigma_z(l_4).
\nonumber\\
\label{eq:HZ2}
\end{eqnarray}
Here, the first (second) sum runs over all links (plaquettes) on the 2D lattice, and open boundary conditions are assumed. A plaquette is defined as the product of four gauge links staring from the lower-left corner and moving counterclockwise, see the upper right panel of Fig.~\ref{fig:z2ising}. The Hamiltonian in Eq.~(\ref{eq:HZ2}) remains invariant under a local gauge transformation which flips the sign of $\sigma_z$ on links sharing site $n$, but does not affect $\sigma_x$ on links sharing the same site. The Gauss's law corresponding to this symmetry defines the physical sector of the theory, namely states for which the eigenvalue of the Gauss's law operator $G(n)=\prod_n \sigma_x(l_n)$ is unity, where $l_n$ denotes all the four links that meet at point $n$.

To establish a duality relation with the 2D Ising model, the gauge invariance can be taken into account to: i) fix the gauge conveniently such that $\sigma_z$ on all links along one of the spatial directions is set to unity, ii) use the operator identity $G(n)=1$ in the physical Hilbert space of the theory to replace $\sigma_x$ along the same space direction as in i) with those along the other direction. These two steps inspire the replacements $\sigma_z(l_1)\sigma_z(l_2)\sigma_z(l_3)\sigma_z(l_4) \to \sigma_x(p)$, and $\prod_{\tilde{l}^\prime \leq \tilde{l}}\sigma_x(\tilde{l}^\prime) \to \sigma_z(p)$ (which is allowed as the new $\{\sigma_x,\sigma_z\}$ set has the same commutation relations as the original set). In the first replacement rule, $p$ denotes the plaquette formed by links $l_1,l_2,l_3,l_4$, and in the second rule, it denotes the plaquette whose left bottom corner is the point at which $\tilde{l}$ starts. The product is over all links prior to and including link $\tilde{l}$, and tilde is used to denote the space dimension for which the gauge remains unfixed. It is now easy to see that in terms of the new spin operators, the Hamiltonian in Eq.~(\ref{eq:HZ2}) can be written as
\begin{eqnarray}
H_{\text{2D Ising}}&=&\lambda\left[ - \sum_p \sigma_x(p) - \frac{1}{\lambda}  \sum_{\langle p,p^\prime \rangle} \sigma_z(p)\sigma_z(p^\prime)\right]
\nonumber\\
&\equiv& -\lambda \sum_{\bm{n}} \sigma_x^{(\bm{n})} - \sum_{\bm{n}}\sum_{j=1,2} \sigma_z^{(\bm{n})} \sigma_z^{(\bm{n}+\hat{\bm{\bm{n}}}_j)},
\label{eq:H2DIsing}
\end{eqnarray}
where in the last line, $\bm{n}$ refers to points on the ``dual'' lattice defined by the center of spatial plaquettes in the original lattice, see Fig.~\ref{fig:z2ising}. $\langle p,p^\prime \rangle$ in the first line denotes the nearest-neighbor plaquettes. For further detail on the expected phase diagram of the theories at different coupling regimes, see e.g., Ref.~\cite{kogut1979introduction}.

The duality between Eq.~(\ref{eq:HZ2}) and (\ref{eq:H2DIsing}) allows to simulate the dynamics of a $Z_2$ LGT in 2+1D using a chain of ions in 1D whose interactions are tailored to correspond to the Ising Hamiltonian, as discussed in the previous example of this section. The correspondence between the original 2D lattice, the dual lattice, and the chain of ions is depicted in Fig.~\ref{fig:z2ising}. Engineering the nearest-neighbor $\sigma_z \otimes \sigma_z$ interactions was detailed in Sec.~\ref{sec:ionphysics}, and the additional global transverse magnetic field can be easily introduced by performing single-qubit rotations, with an angle determined by the coupling $\lambda$ in the original theory.

\
\

\section{Conclusion and outlook
\label{sec:conclusion}}
\noindent
In this paper, we took on the question of how to best leverage the current technologies in ion-trap analog quantum simulators to engineer the Hamiltonian of gauge field theories. Towards this goal, gauge theories that can be experimentally realized in such platforms in the near future are enumerated, and are shown to be amenable to a particular quantum simulation scheme devised in this work. The highlights of the scheme presented, and its promising applications, can be summarized as:
\begin{itemize}
\item[$\rhd$]{$N$ sets of laser beams are used to address individual ions in a 1D chain. With the addition of three global laser beams, the Hamiltonian of a Heisenberg model can be engineered. Certain orientations and frequencies of the beams compared with each other (see Fig~\ref{fig:lasers}) allow $\sigma_x^{(i)}\otimes\sigma_x^{(j)}$, $\sigma_y^{(i)}\otimes\sigma_y^{(j)}$, and $\sigma_z^{(i)}\otimes\sigma_z^{(j)}$ spin-spin interactions to be generated with negligible couplings among different Raman processes. Each set of lasers couples to one set of normal modes of motion (two transverse and one axial), allowing arbitrary spin-spin couplings to be engineered. Our scheme is inspired by that presented in Ref.~\cite{porras2004effective} but does not require an asymmetric trap in the transverse directions, as long as one is interested in a Heisenberg XYZ and XXZ models (see the example of the Schwinger Model in Sec.~\ref{sec:schwinger}).}
\item[$\rhd$]{The experimental scheme of this work offers the capability of engineering a range of interesting dynamics with a single beatnote frequency for each set of the lasers, denoted as $\mu_L$ with $L=I,II,III$, but with tunable phases and with Rabi frequencies $\Omega_L^{(i)}$ at the location of each ion. Moreover, introducing a frequency control to the system, as is common in the digital ion-trap platforms, allows arbitrary spin-spin Hamiltonians to be engineered with unprecedented accuracy.}
\item[$\rhd$]{The frequency control allows an effective local magnetic field to be engineered via asymmetrically shifting the frequency of the red- and blue-detuned Raman beams, eliminating the need for introducing another $N$ laser beams to induce local Stark shifts on the ions.}
\item[$\rhd$]{Engineering an arbitrary Heisenberg Hamiltonian is enabled in this work by a thorough optimization procedure that minimizes the contributions arising from unwanted couplings to phonon excitations, contributions that drive the dynamics away from the effective spin-spin Hamiltonians. This is a crucial requirement for a reliable quantum simulation that is addressed for the first time in this work. The purely spin formulation of the lattice Schwinger model exists, and corresponds to a Heisenberg XXZ model with both short and long-range interactions, and with an effective local magnetic field. The optimization procedure described above was applied to this example with $N=8$, and can be scaled straightforwardly to any number of ions.}
\item[$\rhd$]{In this work, equal-size nearest-neighbor couplings along the spin axes $\hat{x}$ and $\hat{y}$ are achieved through coupling to transverse and axial modes of the motion, respectively, eliminating any significant undesired coupling between the two resulting interacting Hamiltonians in the evolution given the Raman-beam detunings required. This feature does not demand the use of a strong effective magnetic field to induce such nearest-neighbor interactions~\cite{richerme2014non,jurcevic2014quasiparticle,wall2017boson}, with its known limitations~\cite{kiely2018relationship}. Although it may be challenging to implement such a scheme in larger chains of ions with low axial normal-mode frequencies, ideas such as that proposed in Ref.~\cite{brown2011coupled} may allow a scalable scheme in future investigations.
}
\item[$\rhd$]{Another feature of the proposed scheme is a high degree of flexibility in tuning the spin-spin interaction couplings of arbitrary forms along each axis of the qubit independently. This feature, which for example is not offered in single Molmer-Sorenson schemes~\cite{bermudez2017long}, is shown to be particularly useful for engineering the Hamiltonians of gauge theories considered in this work.
}
\item[$\rhd$]{The high level of control allows quantum simulation of models in higher dimensions. Two interesting examples of lattice gauge theories presented in this work (see Sec.~\ref{sec:higherd}) are Abelian Chern-Simons theory coupled to matter, and a $Z_2$ pure gauge theory, both in 2+1D, whose dynamic can be mapped to a planar Ising model with nearest-neighbor interactions. Such capability opens up the possibility of analog quantum simulations of systems beyond what has been possible to date.}
\end{itemize}

A few directions can be recognized as natural extensions of the ideas presented in this paper. These include:
\begin{itemize}
\item[$\lhd$]{There are a range of methods that lead to a truncated angular-momentum representation of the gauge degrees of freedom in LGTs, such as the quantum link models~\cite{Wiese:2013uua,Wiese:2014rla,Rico:2018pas}, or the use of a tensor-network construction in Abelian gauge theories coupled to matter~\cite{Rico:2018pas,Bazavov:2014lla}. With the manipulation of a larger number of internal levels of the ions, the approach advocated in this paper can be applied to engineer interactions of spin systems with $s \geq \frac{1}{2}$. An experimental realization of a spin Hamiltonian with $s=1$ is presented in Ref.~\cite{senko2015realization}, and can be extended to allow quantum simulation of select gauge theories in spin-1 representations.}
\item[$\lhd$]{For a wide range of phenomenologically-interesting lattice gauge theories for which a purely spin representation does not exist, it is essential to extend the toolkit of ion-trap analog simulation to leverage the control over phononic degrees of freedom. This will require further technological  advancement on the experimental front, as well as new proposals for engineering gauge and gauge-matter interactions in a highly controlled spin-phonon system.}
\end{itemize}
%

\section*{Acknowledgments}
\noindent
We are grateful to Jiehang Zhang for his encouragement during the early stages of this interdisciplinary collaboration. We acknowledge valuable discussions with Norbert Linke and Yannick Maurice. ZD is supported in part by the Maryland Center for Fundamental Physics, University of Maryland, College Park. ZD and ASh are supported in part by the U.S. Department of Energy (DOE), Office of Science, Office of Advanced Scientific Computing Research (ASCR) Quantum Computing Application Teams program, under fieldwork proposal number ERKJ347. ASh is further supported by the National Science Foundation (NSF) through the Bridge to the Doctorate Fellowship. MH and ASe are supported by the NSF's Physics Frontier Center at the Joint Quantum Institute (JQI), and by the Air Force Office of Scientific Research, Multidisciplinary University Research Initiative (MURI). CM and GP are supported by the U.S. DOE Basic Energy Sciences (BES) Quantum Computing in Chemical and Material Sciences Program, by the U.S. DOE High-Energy-Physics (HEP) Quantum Information Science Enabled Discovery (QuantISED) Program, by the Army Research Office (ARO) MURI on Modular Quantum Circuits, and by the NSF's Physics Frontier Center at the JQI.

\
\

\appendix
\section{Experimental specifications of the trapped ion system considered for examples of this work
\label{app:Yb}}
\begin{table*}[t!]
\includegraphics[scale=0.635]{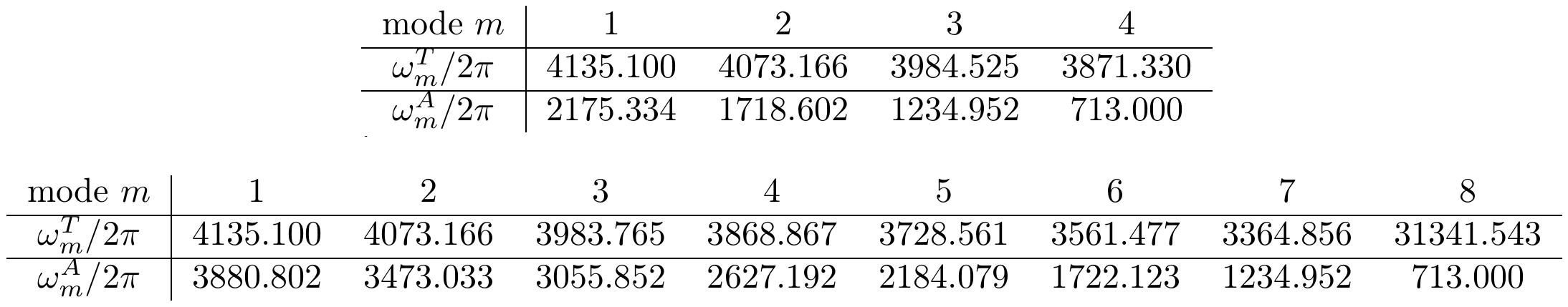}
\caption[.]{Transverse normal modes of the motion of 4 (upper table) and 8 (lower table) ions in the Paul trap considered in this work. Frequencies are in kHz.}
\label{tab:normalmodes}
\end{table*}
\noindent
In order to provide explicit protocols in the examples provided in Sec.~\ref{sec:schwinger} and Appendix~\ref{app:Rabi}, the ion-trap system that is considered is assumed to share similar features as those realized in Refs.~\cite{zhang2017observation,hess2017non,neyenhuis2017observation}. Nonetheless, the general procedure for obtaining these protocols can be identically applied to systems containing other species of ions, and exhibiting different laser characteristics. 

Consider $N$ $^{171}\text{Yb}^+$ ions confined in a radio-frequency Paul trap~\cite{paul1990electromagnetic}. The ``qubit'' in this system has been commonly encoded in a magnetically-insensitive clock state of $^{171}\text{Yb}^+$. However, for the quantum simulations of the gauge theories considered in this study, magnetically-sensitive hyperfine levels $| F=0, m_F=0 \rangle$ and $| F=1, m_F=-1 \rangle$ will be needed, see Fig.~\ref{fig:yblvls}. The former (latter) level corresponds to $s_z=-\frac{1}{2}~(\frac{1}{2})$ component of a quasi-spin operator. These are split in energy by a corresponding frequency $\nu_0\equiv{\omega_0}/{2\pi}=12.642819~\text{GHz}+310.8B_0^2~\text{Hz}/\text{G}^2$, where $B_0$ denotes an external magnetic field~\cite{gill2011should}. Highly efficient state initialization and readout are performed using a laser tuned to $369.5$~nm, which strongly couples the ground $^2S_{1/2}$ and excited $^2P_{1/2}$ states.

For the Paul trap considered in the proposals of this work, $\nu^A= 0.713$ MHz and $\nu^T=4.1351$ MHz, where $\nu^A$ and $\nu^T$ are the axial and transverse frequencies of the confining potential, respectively. The axial and transverse normal-mode frequencies in such a trap are tabulated in Table~\ref{tab:normalmodes} for $N=4$ and $N=8$. Finally, to achieve the values of Lamb-Dicke parameters used in the examples presented in Sec.~\ref{sec:schwinger} and Appendix~\ref{app:N8plots}, the lasers are aligned such that: $\xi=0.6960$ and $\chi=0.1767$, where $\xi$ and $\chi$ are introduced in the caption of Fig.~\ref{fig:lasers}. As a result, the angles between the individual beams and the three global beams $(I)$, $(II)$, and $(III)$ are $88.21^{\circ}$, $20.36^{\circ}$, and $88.21^{\circ}$, respectively.

\section{Tuning spin-dependent forces for the proposed scheme of this work
\label{app:Rabi}}
\noindent
The Hamiltonian $\widetilde{H}_{III}$ in Eq.~(\ref{eq:HintIII}) is proportional to the operator $\alpha_0 \mathbb{I}^{(i)}+\alpha_3 \sigma_z^{(i)}$. As was derived in Sec.~\ref{sec:ionphysics}, the effective spin-spin interaction $H^{(zz)}$ arise from $[\widetilde{H}_{III},\widetilde{H}_{III}]$ commutation at $\mathcal{O}(\eta^2)$ in the Magnus expansion of the time-evolution operator. When $\alpha_0 \neq 0$, this commutation creates an effective $\sigma_z^{(i)}$ Hamiltonian with a strength twice that of the effective $\sigma_z^{(i)} \otimes \sigma_z^{(i)}$ Hamiltonian. Such a bias magnetic field introduces a significant error to the desired evolution. Any attempt to null out such a local magnetic field with additional sets of lasers will cause further nonzero commutations with the $\widetilde{H}_I$ and $\widetilde{H}_{II}$ Hamiltonians, that are generally non-negligible given the strength of the bias magnetic field.\footnote{Such a bias magnetic field term is discussed in Ref.~\cite{schneider2012experimental}.} It is therefore important to investigate solutions that eliminate the term proportional to $\alpha_0$ in the native Hamiltonian in Eq.~(\ref{eq:HintIII}). One such solution relies on tuning the polarizations and detuning of the Raman beams used to produce the $\widetilde{H}_{III}$ Hamiltonian such that the spin-dependent force acting on the state $\ket{\uparrow}$ is negative to that on the state $\ket{\downarrow}$: $F_{\uparrow}=-F_{\downarrow}$. This then sets $\alpha_0=0$, which is the choice used in our proposal in Sec.~\ref{sec:ionphysics}. To demonstrate this solution, we consider the example of $^{171} \rm{Yb}^+$, however, the same approach can be taken to find schemes that work for other ion traps as well.

\begin{figure*}[t!]
\includegraphics[scale=0.3925]{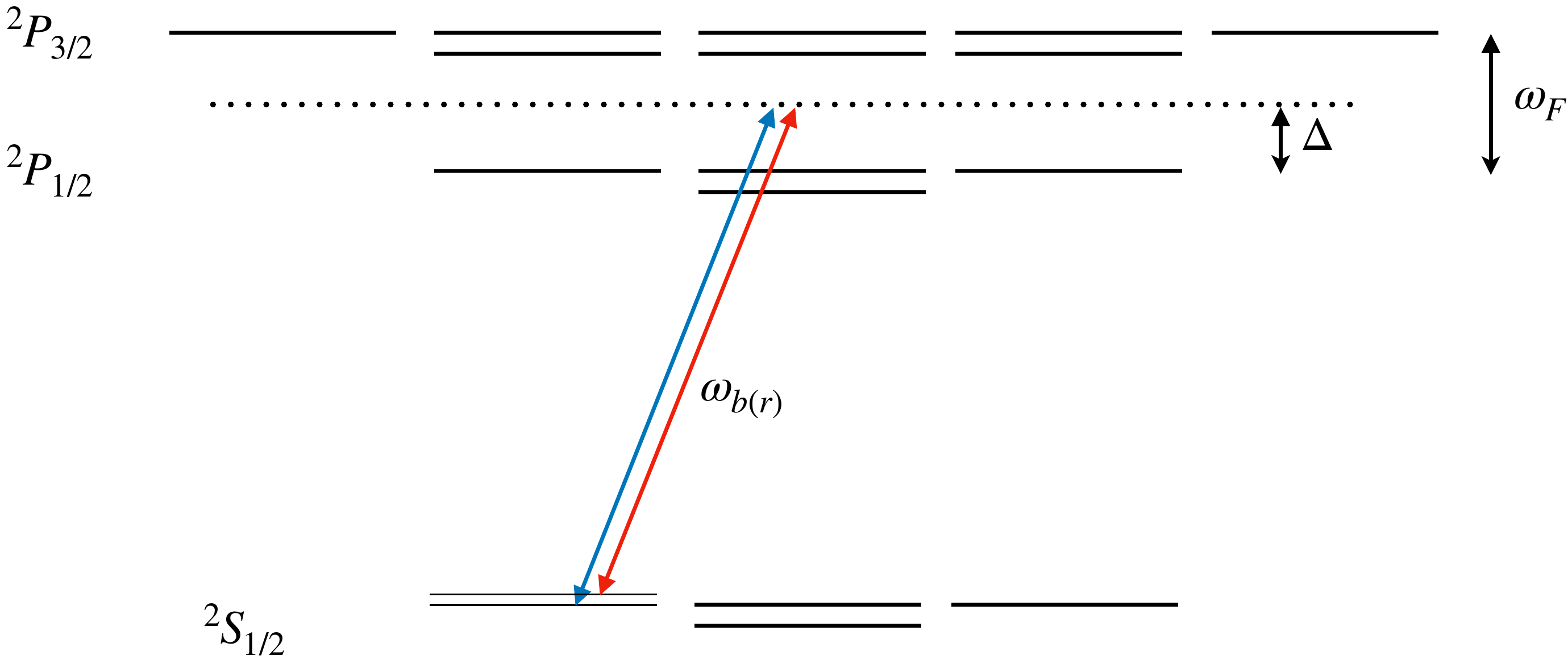}
\caption{The level diagram of $^{171} \rm{Yb}^+$ relevant to the scheme presented in this Appendix.}
\label{fig:yblvls}
\end{figure*}

As mentioned in Appendix~\ref{app:Yb}, the qubit is encoded in the magnetically-sensitive $\ket{\uparrow} \equiv \ket{F=0,m_F = 0}$ and  $\ket{\downarrow} \equiv \ket{F=1,m_F = -1}$ hyperfine $^2 S_{1/2}$ states of $^{171} \rm{Yb}^+$. Consider a set of Raman beams with frequencies $\omega_{r}$ and $\omega_{b}$, detuned from $^2 P_{1/2}$ manifold by $\Delta$. In order to produce a spin-dependent force as discussed in Sec.~\ref{sec:ionphysics}, the beams have to be detuned from each other by the motional mode's frequency $\omega_m$, that is $\Delta\omega=\omega_{b}-\omega_{r}= \omega_m$, see Fig.~\ref{fig:yblvls}. In order to find appropriate polarizations and detuning that allow a pure $\sigma_z$ Hamiltonian, three quantities must be calculated in this scheme: i) the Stark shift induced by red and blue lasers in the Raman pair, ii) the spontaneous emission rate from excited states, and finally iii) the spin-dependent force on the qubit. (iii) must be studied to deduce the conditions under which $F_{\uparrow}=-F_{\downarrow}$, while at the same time (i) must be ensured to vanish, and (ii) must be minimized.

Let us denote the polarization of each beam by $\hat{\bm{\epsilon}}_{r} = r_{-} \hat{\sigma}_{-} + r_0 \hat{\pi} + r_{+} \hat{\sigma}_{+}$ and $\hat{\bm{\epsilon}}_{b} = b_{-} \hat{\sigma}_{-} + b_0 \hat{\pi} + b_{+} \hat{\sigma}_{+}$, where $|r_{-}|^2+ |r_{0}|^2+|r_{+}|^2 =|b_{-}|^2+ |b_{0}|^2+|b_{+}|^2= 1$. In calculating these quantities, matrix elements in the form $\bra{\alpha' F' m'_F} \bm{d}\cdot\hat{\bm{\epsilon}}\ket{\alpha F m_F}$ need to be evaluated, where $\bm{d}$ is the electric dipole operator, and $\alpha$ represents all other quantum numbers of the state besides the total spin $F$ (nuclear spin added to electron's total angular momentum) and its component along the quantization axis, $m_F$. Such a matrix element can be evaluated using~\cite{mizrahi2013ultrafast}
\begin{widetext}
\begin{eqnarray}
\bra{\alpha' F' m'_F} \bm{d}.\hat{\bm{\epsilon}}\ket{\alpha F m_F} &=&
(-1)^{J'+I-m'_F} \sqrt{(2F+1)(2F'+1)}
\begin{Bmatrix}
J' & F' & I \\ 
F & J & 1
\end{Bmatrix}
\begin{pmatrix}
F & 1 & F' \\
m_F & q & -m'_F
\end{pmatrix}
\langle \alpha' J' || d || \alpha J \rangle.
\label{eq:matelem}
\end{eqnarray}
\end{widetext}
Here, $q = -1$ for the $\hat{\sigma}_-$-polarized light, $q = 0$ for the $\hat{\pi}$-polarized light, and $q = 1$ for the $\hat{\sigma}_+$-polarized light. $I$ and $J$ denote the total nuclear spin and the electron's total angular momentum, respectively. ``$()$" corresponds Wigner's $3j$ symbol while ``$\{\}$" corresponds to Wigner's $6j$ symbols. The reduced matrix element $\langle \alpha' J' || d || \alpha J \rangle$ is related to the spontaneous emission rate $\gamma$ between states with $J$ and $J^\prime$ quantum numbers for an atom coupled to free space:
\begin{eqnarray}
|\langle \alpha' J' || d || \alpha J \rangle|^2 =  c_0 (2J'+1) \gamma,
\end{eqnarray}
where $c_0$ is a number that depends on the transitions. For simplicity, in the following we assume that the $^2 P_{1/2}$ and the $^2 P_{3/2}$ states have the same $c_0$ and $\gamma$.

\
\

\noindent
\emph{i) Stark shift:} In the limit where $\Delta\gg\gamma$, the Stark shift for $\ket{m_S} = \ket{\uparrow}, \ket{\downarrow}$ is given by~\cite{wineland2003quantum}
\begin{eqnarray}
\delta_{\text{Stark}}(m_S)=\frac{1}{4} \sum_{j=r,b}\sum_{i} \frac{|\bra{m_S} \bm{d}\cdot\hat{\bm{\epsilon}}_jE_j\ket{i}|^2}{\Delta_i},
\end{eqnarray}
where $\Delta_i$ is the detuning from the states that are virtually occupied, and $E_j$ is the electric-field amplitude. Using Eq.~\eqref{eq:matelem}, the net Stark shift is found to be
\begin{eqnarray}
&& \delta_{\text{Stark}}(\uparrow)-\delta_{\text{Stark}}(\downarrow)=\frac{c_0 \gamma \omega_F  }{12   ^2 \Delta  (\Delta -\omega_F)}
\nonumber\\
&& \hspace{1.75 cm} \times \left(\left| b_-\right|^2+\left| r_-\right|^2-\left| b_+\right|^2-\left| r_+\right|^2 \right).
\label{eq:starkshift}
\end{eqnarray}
As is evident, by choosing $\left| b_-\right|^2+\left| r_-\right|^2=\left| b_+\right|^2+\left| r_+\right|^2 $, the net shift can be set to zero. 

\
\

\noindent
\emph{ii) Spontaneous emission:} The spontaneous emission rate can be evaluated using~\cite{wineland2003quantum}
\begin{eqnarray}
R_{\rm{SE}} &=& 
\frac{1}{4 } \sum_i \sum_{j=r,b} 
\sum_{m_S = \downarrow,\uparrow}  \frac{P_{m_S}\gamma_i |\bra{m_S}\bm{d}\cdot\hat{\bm{\epsilon}}_jE_j\ket{i}|^2}{\Delta_i^2},
\nonumber\\
\end{eqnarray}
where $P_{m_S}$ is the probability of being in the $m_S$ ground state. Under the constraint that sets Eq.~\eqref{eq:starkshift} to zero, one finds that
\begin{eqnarray}
R_{\rm{SE}} = \frac{c_0 \gamma^2(2+|r_0|^2+|b_0|^2)}{12\sqrt{(1+|r_0|^2)(1+|b_0|^2)}} \left[\frac{1}{\Delta^2}+ \frac{2}{(\Delta-\omega_F)^2}\right].
\nonumber\\
\label{eq:spontemiss}
\end{eqnarray}
As is seen, with the choice $\Delta = (\sqrt{2}-1)\omega_F$ one is close to a local minimum of the spontaneous emission rate. 

\
\

\noindent
\emph{iii) Spin-dependent force:} Finally, the spin-dependent force can be found by considering the resonant two-photon Raman Rabi rate~\cite{wineland2003quantum}
\begin{equation}
\Omega(m_S) = \frac{e^{i(\varphi_b-\varphi_r)}}{4} \sum_i \frac{\bra{m_S}\bm{d}\cdot\hat{\bm{\epsilon}}_rE_r\ket{i}\bra{i}\bm{d}\cdot\hat{\bm{\epsilon}}_bE_b\ket{m_S}}{\Delta_i},
\end{equation}
where $\varphi_r$ and $\varphi_b$ are the phases of the red- and blue-detuned beams, respectively. With $\Delta\varphi \equiv \varphi_b-\varphi_r=0$ and $\Delta= (\sqrt{2}-1)\omega_F$, one find that
\begin{align}
\Omega(\downarrow) &= \frac{-\gamma (b_0 r_0^*+b_- r_-^*+b_+ r_+^*)}{12 \omega_F }, \\
\Omega(\uparrow) &= \frac{\gamma (- 2 b_0 r_0^*+(2+3\sqrt{2})b_+ r_+^*-3(2+\sqrt{2}) b_- r_-^*)}{24 \omega_F }.
\end{align}
In order to satisfy the condition $\Omega(\downarrow)=-\Omega(\uparrow)$ or in turn $F_{\uparrow}=-F_{\downarrow}$,\footnote{Note that the spin-dependent force is related to the Rabi frequency via $F_{m_S} = \Delta k~\Omega(m_S)$.} a choice for the polarization vectors is
\begin{align}
\hat{\bm{\epsilon}}_{b} &= \frac{3}{2-\sqrt{2}} (-1,\sqrt{2+\frac{3}{\sqrt{2}}},1),\\
\hat{\bm{\epsilon}}_{r} &=  \frac{3}{2-\sqrt{2}} (1,\sqrt{2+\frac{3}{\sqrt{2}}},1).
\end{align}
Of course, these analytical solutions rely on the approximations that were made throughout these calculations, such as equal spontaneous emission rate from all the excited states considered. When precise values of the physical parameters in the system are input, the optimal values for the parameters can still be evaluated numerically using the formalism outlined. See also Ref.~\cite{britton2012engineered} for a similar approach in achieving the condition $F_{\uparrow}=-F_{\downarrow}$.

\
\

\section{Details of the laser-ion evolution operator
\label{app:H}}
\noindent
In this Appendix, the explicit forms of the functions appeared in Eqs.~(\ref{eq:UAll}-\ref{eq:phiZ}) of the main text will be provided. The following frequency parameters are used:
\begin{eqnarray}
&&\Delta_m^T \equiv \mu_I+\omega_m^T,~\delta_m^T \equiv \mu_I-\omega_m^T,
\\
&&\Delta_m^A \equiv \mu_{II}+\omega_m^A,~\delta_m^A \equiv \mu_{II}-\omega_m^A,
\\
&&\widetilde{\Delta}_m^T \equiv \mu_{III}+\omega_m^T,~\widetilde{\delta}_m^T \equiv \mu_{III}-\omega_m^T,
\end{eqnarray}
while the rest of the parameters/functions are already defined in Sec.~\ref{sec:ionphysics}.

\begin{widetext}
\begin{eqnarray}
\label{eq:alphaxdef}
&&\alpha^{(x)}_{i,m}(t)=\frac{\eta_{I,m}^{(i)}\Omega_{I}^{(i)}}{2}\left[
\int_0^t dt_1 \left(e^{i\Delta_m^T t_1}-e^{-i\delta_m^T t_1} \right)+\frac{iB^{(i)}}{2} \int_0^t dt_2 \int_0^{t_2} dt_1 \left[\left(e^{i\Delta_m^A t_1}+e^{-i\delta_m^A t_1} \right) - t_1 \leftrightarrow t_2 \right] \right],
\\
\label{eq:alphaydef}
&&\alpha^{(y)}_{i,m}(t)=\frac{i\eta_{II,m}^{(i)}\Omega_{II}^{(i)}}{2}\left[
\int_0^t dt_1 \left(e^{i\Delta_m^A t_1}+e^{-i\delta_m^A t_1} \right)-\frac{iB^{(i)}}{2} \int_0^t dt_2 \int_0^{t_2} dt_1 \left[\left(e^{i\Delta_m^T t_1}-e^{-i\delta_m^T t_1} \right) - t_1 \leftrightarrow t_2 \right] \right],
\\
%
&&\alpha^{(z)}_{i,m}(t)=\frac{\eta_{III,m}^{(i)}\Omega_{III}^{(i)}}{2}\int_0^t dt_1 \left(e^{i\widetilde{\Delta}_m^T t_1}-e^{-i\widetilde{\delta}_m^T t_1} \right).
\label{eq:alphazdef}
\end{eqnarray}
\begin{eqnarray}
\label{eq:betaxdef}
&&\beta^{(x)}_{i,m,n}(t)=\frac{\eta_{II,m}^{(i)}\eta_{III,n}^{(i)}\Omega_{II}^{(i)}\Omega_{III}^{(i)}}{2} \int_0^t dt_2 \int_0^{t_2} dt_1 \left[\left(e^{i\Delta_m^A t_2}+e^{-i\delta_m^A t_2} \right)\left(e^{i\widetilde{\Delta}_n^T t_1}-e^{-i\widetilde{\delta}_n^T t_1} \right) - t_1 \leftrightarrow t_2\right],
\\
\label{eq:betaydef}
&&\beta^{(y)}_{i,m,n}(t)=-\frac{i\eta_{I,m}^{(i)}\eta_{III,n}^{(i)}\Omega_{I}^{(i)}\Omega_{III}^{(i)}}{4} \int_0^t dt_2 \int_0^{t_2} dt_1 \left[\left(e^{i\Delta_m^T t_2}-e^{-i\delta_m^T t_2} \right)\left(e^{i\widetilde{\Delta}_n^T t_1}-e^{-i\widetilde{\delta}_n^T t_1} \right) - t_1 \leftrightarrow t_2\right],
\\
&&\beta^{(z)}_{i,m,n}(t)=\frac{\eta_{I,m}^{(i)}\eta_{II,n}^{(i)}\Omega_{I}^{(i)}\Omega_{II}^{(i)}}{4} \int_0^t dt_2 \int_0^{t_2} dt_1 \left[\left(e^{i\Delta_m^T t_2}-e^{-i\delta_m^T t_2} \right)\left(e^{i\Delta_n^A t_1}+e^{-i\delta_n^A t_1} \right) - t_1 \leftrightarrow t_2\right].
\label{eq:betazdef}
\end{eqnarray}
\begin{eqnarray}
\gamma^{(z)}_{i}(t) = \frac{iB^{(i)}}{4} \int_0^{t} dt_1.
\label{eq:gammadef}
\end{eqnarray}
\begin{eqnarray}
\label{eq:chixdef}
\chi^{(x)}_{i,j}(t) &=& \sum_{m=1}^N\frac{\eta_{I,m}^{(i)}\eta_{I,m}^{(j)}\Omega_{I}^{(i)}\Omega_{I}^{(j)}}{8}  \int_0^t dt_2 \int_0^{t_2} dt_1 \left[\left(e^{i\Delta_m^T t_2}-e^{-i\delta_m^T t_2} \right) \left(e^{i\Delta_m^T t_1}-e^{-i\delta_m^T t_1} \right) \right],
\\
\label{eq:chiydef}
\chi^{(y)}_{i,j}(t) &=& - \sum_{m=1}^N\frac{\eta_{II,m}^{(i)}\eta_{II,m}^{(j)}\Omega_{II}^{(i)}\Omega_{II}^{(j)}}{8}  \int_0^t dt_2 \int_0^{t_2} dt_1 \left[\left(e^{i\Delta_m^A t_2}+e^{-i\delta_m^A t_2} \right) \left(e^{i\Delta_m^A t_1}+e^{-i\delta_m^A t_1} \right) \right],
\\
\chi^{(z)}_{i,j}(t) &=&  \sum_{m=1}^N\frac{\eta_{III,m}^{(i)}\eta_{III,m}^{(j)}\Omega_{III}^{(i)}\Omega_{III}^{(j)}}{8}  \int_0^t dt_2 \int_0^{t_2} dt_1 \left[\left(e^{i\widetilde{\Delta}_m^T t_2}-e^{-i\widetilde{\delta}_m^T t_2} \right) \left(e^{i\widetilde{\Delta}_m^T t_1}-e^{-i\widetilde{\delta}_m^T t_1} \right) \right].
\label{eq:chizdef}
\end{eqnarray}
\end{widetext}

%
\begin{figure*}[h!]
\includegraphics[scale=0.65]{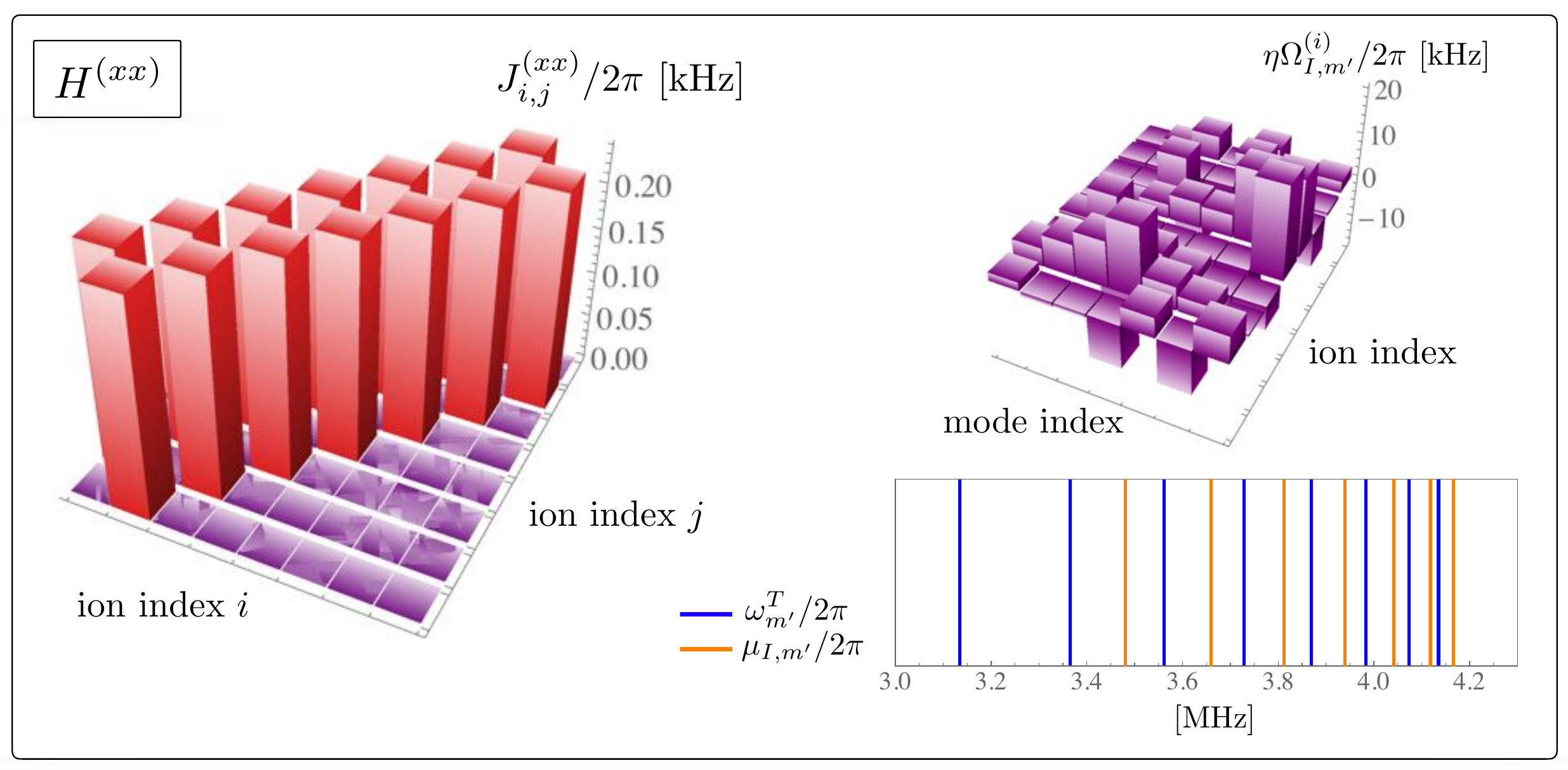}
\caption[.]{The effective spin-spin coupling matrix $J^{(xx)}$ in Eq.~(\ref{eq:JXXdef}) resulting from multiple pairs of Raman beams addressing $N=8$ individual ions at the Rabi frequency $\Omega_{I,m^\prime}^{(i)}$, where $i=1,\cdots,8$ and $m^\prime=1,\cdots,7$. The pairs of beams addressed at ion $i$ are detuned from the transverse COM mode by 7 different frequencies, $\mu_{I,m^\prime}=\omega_{m^\prime}^T+f_s(\omega_{m^\prime}^T-\omega_{m^\prime+1}^T)$ with $f_s=0.5$, as denoted in the lower-right of the panel. The Lamb-Dicke parameter, $\eta$, multiplying the Rabi frequencies in the figure is $\eta=\sqrt{{(\Delta k_I)^2}/{4\pi M\nu^T}} \approx 0.068$. Here, the $J^{(xx)}$ matrix is tuned to produce $H^{(xx)}$ of the 8 fermion-site  Schwinger model in Eq.~(\ref{eq:HSchwingerSplitI}) with $x=6$. Numerical values associated with this figure are provided in Supplemental Material.}
\label{fig:xxnnfor8ions}
\end{figure*}
\begin{figure*}[h!]
\includegraphics[scale=0.65]{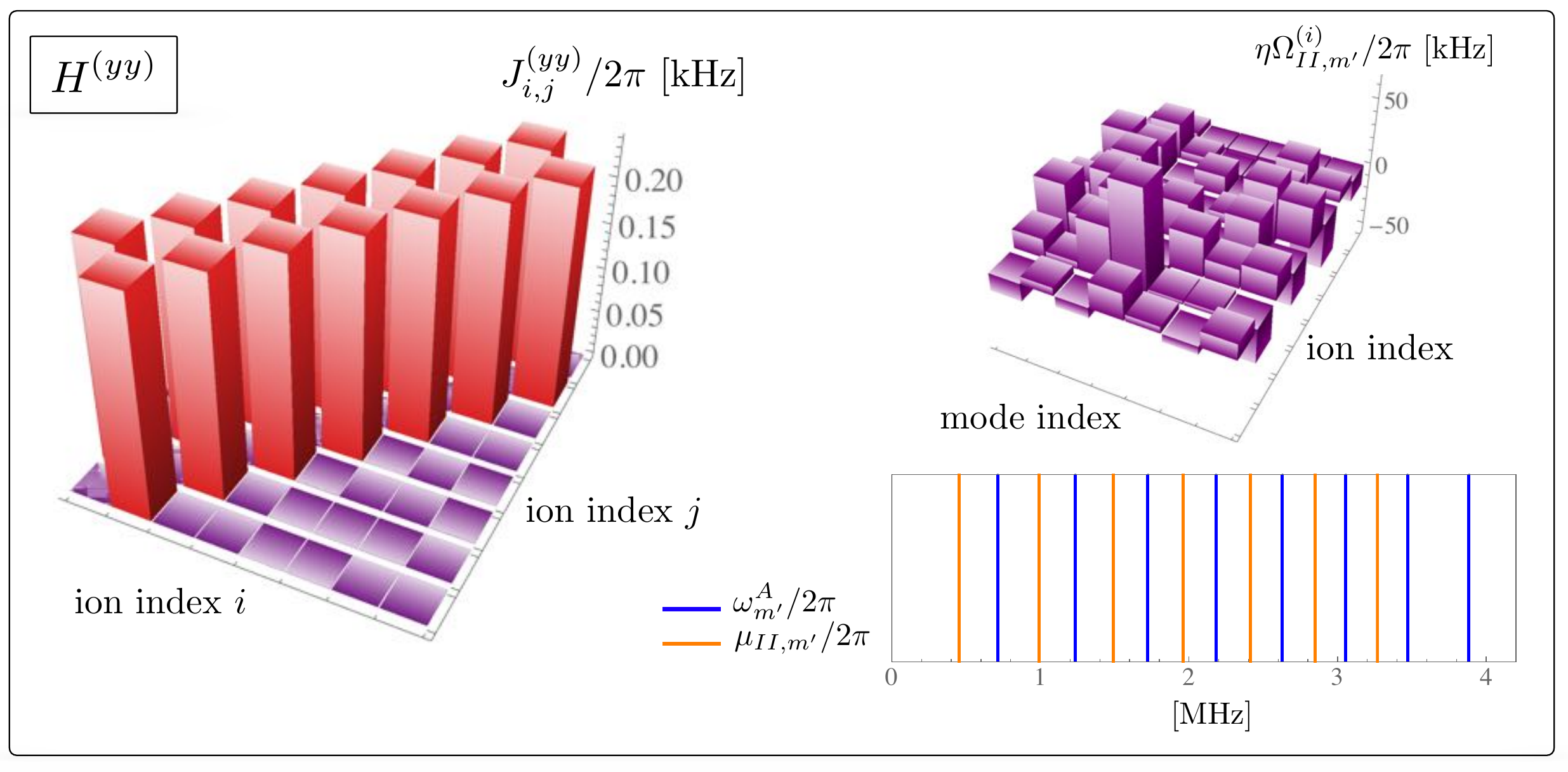}
\caption[.]{The effective spin-spin coupling matrix $J^{(yy)}$ in Eq.~(\ref{eq:JYYdef}) resulting from multiple pairs of Raman beams addressing $N=8$ individual ions at the Rabi frequency $\Omega_{II,m^\prime}^{(i)}$, where $i=1,\cdots,8$ and $m^\prime=1,\cdots,7$. The pairs of beams addressed at ion $i$ are detuned from the axial COM mode by 7 different frequencies, $\mu_{I,N-m^\prime+1}=\omega_{N-m^\prime+1}^A+f_s(\omega_{N-m^\prime}^A-\omega_{N-m^\prime+1}^A)$ with $f_s=-0.5$, as denoted in the lower-right of the panel. The Lamb-Dicke parameter, $\eta$, multiplying the Rabi frequencies in the figure is $\eta=\sqrt{{(\Delta k_{II})^2}/{4\pi M\nu^A}} \approx 0.081$. Here, the $J^{(yy)}$ matrix is tuned to produce $H^{(yy)}$ of the 8 fermion-site  Schwinger model in Eq.~(\ref{eq:HSchwingerSplitII}) with $x=6$. Numerical values associated with this figure are provided in Supplemental Material.}
\label{fig:yynnfor8ions}
\end{figure*}
\begin{figure*}[t!]
\includegraphics[scale=0.65]{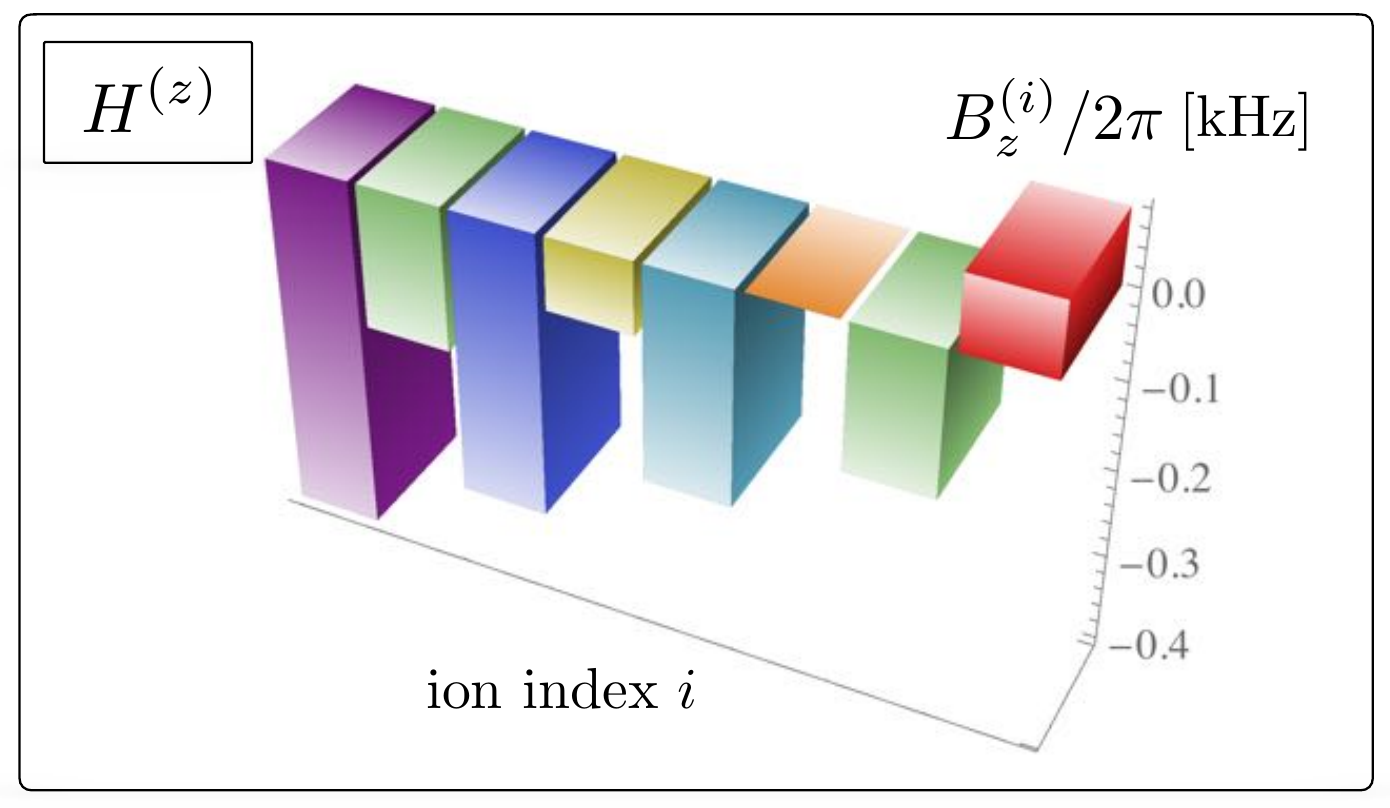}
\caption[.]{The effective magnetic field, $B_z$, that produces the $H^{(z)}$ Hamiltonian of the Schwinger model, Eq.~(\ref{eq:HSchwingerSplitIV}), for $N=8$ and $\mu=1$. Numerical values associated with this figure are provided in Supplemental Material.}
\label{fig:bzfor8ions}
\end{figure*}

\section{Engineered Hamiltonian of the Schwinger model with $N=8$ ions
\label{app:N8plots}}
The multi-frequency, multi-amplitude scheme presented at the end of Sec.~\ref{sec:schwinger} describes the engineering of the long-range Hamiltonian of the Schwinger model in the 8 fermion-site theory, see Fig.~\ref{fig:zzlrfor8ions}. The same optimization procedure can be adopted to engineer the nearest-neighbor Hamiltonians in the same theory using sets of laser beams that address transverse (for $H^{(xx)}$) and axial (for $H^{(yy)}$) normal modes of motion. The associated results, as well as the required effective magnetic field that produces $H^{(z)}$, are depicted in Figs.~\ref{fig:xxnnfor8ions}-\ref{fig:bzfor8ions} of this appendix. Associated numerical values are presented in Supplemental Material.

\section{Numerical evaluation of lasers-ions evolution
\label{app:numerics}}
\begin{figure*}[t!]
\includegraphics[scale=0.655]{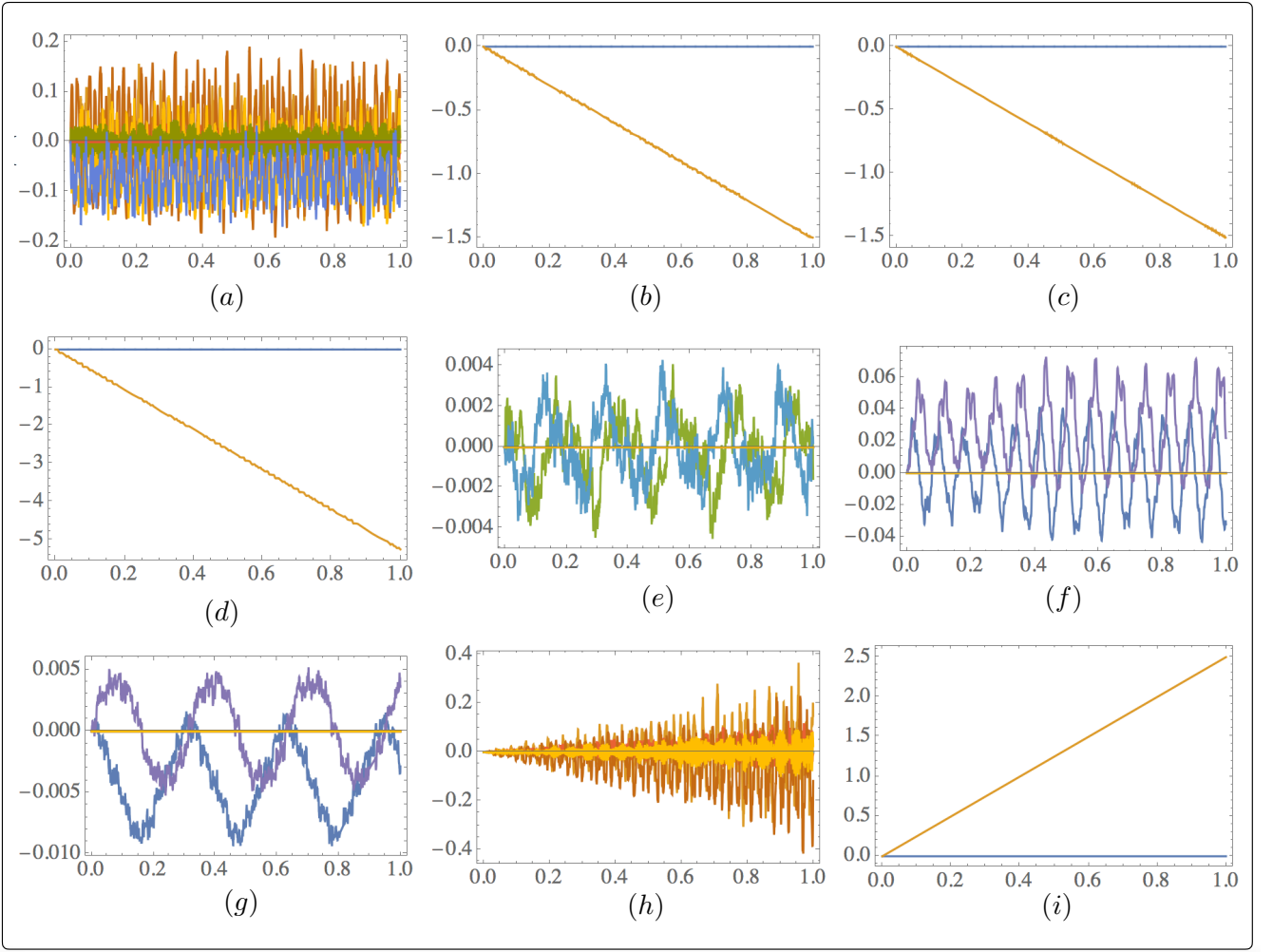}
\caption[.]{Contributions to the exponent of the full laser-ion evolution operator up to and including $\mathcal{O}(\eta^2,\eta B)$ for laser parameters found in the multi-frequency, multi-amplitude scheme in Sec.~\ref{sec:schwinger} to engineer the 8 fermion-site Schwinger Hamiltonian with $x=6$ and $\mu=1$. The quantities plotted are enumerated in this Appendix and are dimensionless. The horizontal axis is time in ms. The plots shown correspond to the evolution of the first ion in the chain. The results for the rest of the ions can be found in Supplemental Material.}
\label{fig:allcontbion1for8ions}
\end{figure*}

In order to confirm that the evolution of laser-ion systems in the scheme proposed in this work follows that of a Heisenberg spin model with a magnetic field, the exponent of the full evolution operator up to $\mathcal{O}(\eta^2,\eta B)$ (see Eq.~(\ref{eq:UAll})) can be numerically evaluated for each set of laser beatnote and Rabi frequencies found. Here, we assume that the ions are in their motional ground state, which can be achieved in current ion-trap experiments. The results of this evaluation are plotted, respectively, in Figs.~\ref{fig:allcontbion1for4ions} and \ref{fig:allcontbion1for8ions} for the case of the Schwinger-model parameters with $N=4$ and $N=8$ that were studied in Sec.~\ref{sec:schwinger}. These figures correspond to the evolution of the first ion in the chain and the results for the rest of the ions are included in Supplemental Material. To interpret these plots, note that the quantities that are plotted are contributions to the exponent of the evolution operator as a function of time $t$ in millisecond (ms), and that:
\begin{figure*}[t!]
\includegraphics[scale=0.655]{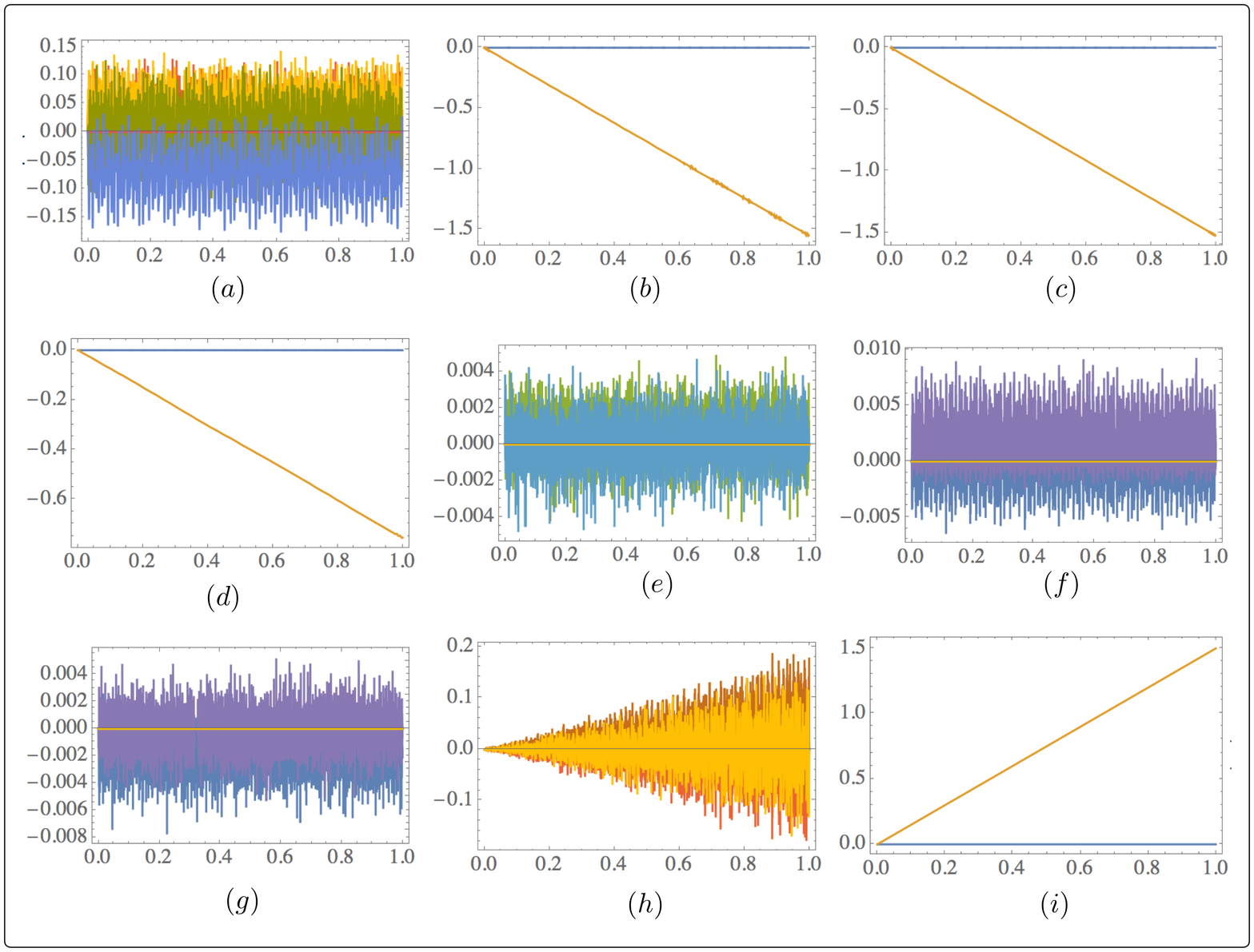}
\caption[.]{Contributions to the exponent of the full laser-ion evolution operator up to and including $\mathcal{O}(\eta^2,\eta B)$ for laser parameters found in the single-frequency, multi-amplitude scheme in Sec.~\ref{sec:schwinger} to engineer the 4 fermion-site Schwinger Hamiltonian with $x=6$ and $\mu=1$. The quantities plotted are enumerated in this Appendix and are dimensionless. The horizontal axis is time in ms. The plots shown correspond to the evolution of the first ion in the chain. The results for the rest of the ions can be found in Supplemental Material.}
\label{fig:allcontbion1for4ions}
\end{figure*}
\begin{itemize}
\item[-] {$(a)$ plots in different colors the real and imaginary parts of all contributions arising from $-i\int_0^t dt_1 H_L^\prime(t_1)$ with $L=I,II,III$ acting on a state with phonon number $n_{\text{ph}}=0$, and ignoring the $\mathcal{O}(1)$ numerical factor arising from spin operators acting on a general spin state. These are referred to as first-order terms, $\mathcal{O}(\eta)$, elsewhere.}
\item[-] {$(b)$, $(c)$, and $(d)$ plot in different colors the real and imaginary parts of all contributions arising from $-\frac{1}{2}\int_0^t dt_2 \int_0^{t_2} dt_1 [\widetilde{H}_I(t_2),\widetilde{H}_I(t_1)]$, $-\frac{1}{2}\int_0^t dt_2 \int_0^{t_2} dt_1 [\widetilde{H}_{II}(t_2),\widetilde{H}_{II}(t_1)]$, and $-\frac{1}{2}\int_0^tdt_2 $ $ \int_0^{t_2} dt_1 [\widetilde{H}_{III}(t_2),\widetilde{H}_{III}(t_1)]$, respectively, acting on a state with $n_{\text{ph}}=0$, and ignoring the $\mathcal{O}(1)$ numerical factor arising from spin operators acting on a general spin state. As is seen, effective $H^{(xx)}$, $H^{(yy)}$, and $H^{(zz)}$ Hamiltonians originate from the imaginary part of these contributions, signified by an almost exact linear dependence in time.}
\item[-] {$(e)$, $(f)$, and $(g)$ plot in different colors the real and imaginary parts of all contributions arising from $-\frac{1}{2}\int_0^t dt_2 \int_0^{t_2} dt_1 ([\widetilde{H}_I(t_2),\widetilde{H}_{II}(t_1)]+[\widetilde{H}_{II}(t_2),\widetilde{H}_I(t_1)])$, $-\frac{1}{2}\int_0^t dt_2 \int_0^{t_2} dt_1 ([\widetilde{H}_I(t_2),\widetilde{H}_{III}$ $(t_1)]+[\widetilde{H}_{III}(t_2),\widetilde{H}_I(t_1)])$, and $-\frac{1}{2}\int_0^t dt_2 \int_0^{t_2} $ $dt_1 ([\widetilde{H}_{II}(t_2),\widetilde{H}_{III}(t_1)]+[\widetilde{H}_{III}(t_2),\widetilde{H}_{II}(t_1)])$, respectively, acting on a state with $n_{\text{ph}}=0$, and ignoring the $\mathcal{O}(1)$ numerical factor arising from spin operators acting on a general spin state. The small contributions observed show that the choice of lasers' detunings in our scheme leads to negligible commutations among the two sets of the lasers.}
\item[-] {$(h)$ plots in different colors the real and imaginary parts of all contributions arising from $-\frac{1}{2}\int_0^t dt_2 \int_0^{t_2} dt_1 ([H_B(t_2),H_L^\prime(t_1)]+[H_L^\prime(t_2),H_B(t_1)])$ for $L=I,II,III$ acting on a state with $n_{\text{ph}}=0$, and ignoring the $\mathcal{O}(1)$ numerical factor arising from spin operators acting on a general spin state. While these contributions are assured to remain a small fraction of the effective magnetic field desired, they are not bounded in time and couple to motional degrees of freedom. As a result, these contributions constitute the largest error to the desired effective-Hamiltonian description that is engineered.}
\item[-] {$(i)$ plots in different colors the real and imaginary parts of the contributions arising from $-i\int_0^t dt_1 H_B(t_1)$ acting on a state with phonon number $n_{\text{ph}}=0$, and ignoring the $\mathcal{O}(1)$ numerical factor arising from spin operators acting on a general spin state. The real part of this contribution corresponds to the desired $H^{(z)}$ Hamiltonian.}
\end{itemize}
Note that in the multi-frequency, multi-amplitude scheme applied to the case of $N=8$, the Hamiltonians in Eqs.~(\ref{eq:HintI}-\ref{eq:HintIII}) must be generalized as described in Sec.~\ref{sec:schwinger} (see discussions after Eq.~(\ref{eq:JijGeneral})). The relation between the contributions enumerated and those given in Eqs.~(\ref{eq:UAll}-\ref{eq:phiZ}) and (\ref{eq:alphaxdef}-\ref{eq:chizdef}) is evident.

\bibliography{bibi.bib}
\end{document}